\documentclass[aps,prd,amsfonts,amssymb, amsmath, showpacs, showkeys, a4paper, nofootinbib, superscriptaddress, 11pt, raggedbottom]{revtex4-2}

\usepackage{braket}
\usepackage{bm}
\usepackage{graphicx}
\usepackage{slashed}
\usepackage{subfig}
\usepackage{mathrsfs}
\usepackage{color}
\usepackage{mathtools}
\usepackage[normalem]{ulem}
\usepackage{simplewick}

\newcommand{\kket}[1]{| #1 \rangle\!\rangle}
\newcommand{\bbra}[1]{\langle\!\langle #1 |}
\newcommand{\bbrakket}[1]{\langle\!\langle #1 \rangle\!\rangle}

\DeclareMathOperator*{\sumint}{%
\mathchoice%
  {\ooalign{$\displaystyle\sum$\cr\hidewidth$\displaystyle\int$\hidewidth\cr}}
  {\ooalign{\raisebox{.14\height}{\scalebox{.7}{$\textstyle\sum$}}\cr\hidewidth$\textstyle\int$\hidewidth\cr}}
  {\ooalign{\raisebox{.2\height}{\scalebox{.6}{$\scriptstyle\sum$}}\cr$\scriptstyle\int$\cr}}
  {\ooalign{\raisebox{.2\height}{\scalebox{.6}{$\scriptstyle\sum$}}\cr$\scriptstyle\int$\cr}}
}

\begin{document}

\title{\Large Density matrices in quantum field theory: Non-Markovianity, path integrals and master equations}

\author{Christian K\"{a}ding}
\email{christian.kaeding@tuwien.ac.at}
\affiliation{Atominstitut, Technische Universit\"at Wien, Stadionallee 2, 1020 Vienna, Austria}

\author{Mario Pitschmann}
\email{mario.pitschmann@tuwien.ac.at}
\affiliation{Atominstitut, Technische Universit\"at Wien, Stadionallee 2, 1020 Vienna, Austria}

\begin{abstract}
Density matrices are powerful mathematical tools for the description of closed and open quantum systems. Recently, methods for the direct computation of density matrix elements in scalar quantum field theory were developed based on thermo field dynamics (TFD) and the Schwinger-Keldysh formalism. In this article, we provide a more detailed discussion of these methods and derive expressions for density matrix elements of closed and open systems. At first, we look at closed systems by discussing general solutions to the Schr\"odinger-like form of the quantum Liouville equations in TFD, showing that the dynamical map is indeed divisible, deriving a path integral-based expression for the density matrix elements in Fock space, and explaining why perturbation theory enables us to use the last even in situations where all initial states in Fock space are occupied. Subsequently, we discuss open systems in the same manner after tracing out environmental degrees of freedom from the solutions for closed systems. We find that, even in a general basis, the dynamical map is not divisible, which renders the dynamics of open systems non-Markovian. Finally, we show how the resulting expressions for open systems can be used to obtain quantum master equations, and comment on the artificiality of time integrals over density matrices that usually appear in many other master equations in the literature but are absent in ours. 
\end{abstract}


\maketitle



\section{Introduction}

Basically, most realistic quantum systems have to be treated as open, which means that they are interacting with some type of environment that impacts the system but whose own dynamics is not fully considered. In practice, this is realized by tracing out the environmental degrees of freedom from the mathematical description of the total system consisting of the subsystems of interest and their environments. Open quantum systems not only find applications in non-relativistic physics, see, e.g.,~Refs.~\cite{Carmichael,Gardiner2004,Walls2008,Aolita2015,Goold2016,Werner2016,Huber2020,Keefe:2024cia,Cavina:2025xnz}, but also in quantum field theory \cite{Calzetta2008,Koksma2010,Koksma2011,Sieberer2016,Marino2016,Baidya2017,Burrage2018,Burrage2019,Banerjee:2020ljo,Nagy2020,Banerjee:2021lqu,Jana2021,Fogedby2022,Kading:2022jjl,Cao:2023syu,Bowen:2024emo,Reyes-Osorio:2024chg,Fahn:2024fgc,Kashiwagi:2024fuy,Salcedo:2024nex,Burrage:2025xac,Farooq:2025pxt}, cosmology \cite{Lombardo1,Lombardo2,Lombardo3,Boyanovsky1,Boyanovsky2,Boyanovsky3,Boyanovsky4,Burgess2015,Hollowood,Shandera:2017qkg,Choudhury:2018rjl,Bohra:2019wxu,Akhtar:2019qdn,Binder2021,Brahma:2021mng,Cao:2022kjn,Brahma2022,Colas:2022hlq,Colas:2022kfu,Colas:2023wxa,Kading:2023mdk,Bhattacharyya:2024duw,Colas:2024xjy,Burgess:2024eng,Salcedo:2024smn,Colas:2024lse,Colas:2024ysu,Brahma:2024yor,Kading:2024jqe,Brahma:2024ycc,Burgess:2024heo,Lau:2024mqm,Takeda:2025cye}, black hole physics \cite{Yu2008,Lombardo2012,Jana2020,Agarwal2020,Kaplanek2020,Burgess2021,Kaplanek2021}, or heavy-ion physics \cite{Brambilla1,Brambilla2,Yao2018,Yao2020,Akamatsu2020,DeJong2020,Yao2021,Brambilla2021,Griend2021,Yao2022}.

Density matrices are frequently used to mathematically describe open quantum systems. Their elements can be obtained by either solving quantum master equations or by directly computing them; see, for example, Refs.~\cite{Kading:2022jjl,Kading:2022hhc}. The open dynamics described by density matrices can be Markovian, i.e.,~without memory effects of the environment, or non-Markovian, which means that changes in the environment at some earlier time may affect the system's dynamics at a later time. An often employed definition of Markovianity is the concept of CP-divisibility \cite{Rivas:2010,Chruscinski:2022hvy}. It means that the dynamical map $\Phi(t,0)$ that brings the density matrix from an initial time $0$ to a later time $t$ is completely positive (CP) and divisible, i.e.,~$\Phi(t,0) = \Phi(t,t')\Phi(t',0)$ for some intermediate time $t'$. However, as it was pointed out in Ref.~\cite{Milz2019}, CP-divisibility is not a sufficient condition for Markovianity. Though, conversely, Markovianity implies divisibility \cite{Milz2019}. Therefore, if divisibility is broken, this indicates non-Markovian dynamics.

Employing earlier results from Refs.~\cite{Burrage2018,Burrage2019}, Ref.~\cite{Kading:2022jjl} recently presented a formalism for directly computing reduced density matrix elements in a Fock basis of scalar quantum field theory, which is based on thermo field dynamics (TFD) \cite{Takahasi:1974zn,Arimitsu:1985ez,Arimitsu:1985xm, Khanna} and the Schwinger-Keldysh formalism \cite{Schwinger,Keldysh}. In Ref.~\cite{Kading:2022hhc}, this formalism was adapted for closed systems. Both Refs.~\cite{Burrage2018, Kading:2022jjl} claim to provide non-Markovian descriptions of open quantum systems, but do not discuss this in any detail. In addition, even though the results of both articles are closely related, as of yet, it has not been investigated how the master equation approach from Ref.~\cite{Burrage2018} can be recovered from the direct computation method in Ref.~\cite{Kading:2022jjl}. 

In this article, we provide an accessible introduction to the methods presented in Refs.~\cite{Kading:2022jjl,Kading:2022hhc} and discuss the open issues mentioned above. For this, we first consider closed systems and discuss solutions to the Schr\"odinger-like form of the quantum Liouville equation in TFD. We do our discussion both for the exact case and, assuming all self-interactions and interactions of the considered subsystems to be sufficiently small, up to second order in perturbation theory. Subsequently, we show that, as it should be, the dynamical map for a general closed system is divisible. We then derive an expression for the total density matrix elements in a momentum basis in Fock space, and explain why it is a powerful tool even when every initial state in Fock space is occupied. After the discussion of closed systems, we consider open quantum systems. In particular, we take the solutions of the Schr\"odinger-like quantum Liouville equation and trace out environmental degrees of freedom. Using the resulting expression, we show that, even for a general basis, the dynamical map is not divisible. After applying the Born approximation in the perturbative case, we find that divisibility is broken by connected correlation functions of the environment and the system-environment interactions. This supports the claims of Refs.~\cite{Burrage2018,Kading:2022jjl} that their methods describe dynamics that are not Markovian. After finding expressions for reduced density matrix elements in a Fock basis, we demonstrate how to obtain quantum master equations from the results of Ref.~\cite{Kading:2022jjl}, which allows us to recover the master equation approach presented in Ref.~\cite{Burrage2018}.


\section{Closed systems}
\label{sec:closedsys}

We begin our investigation by looking at general solutions to the quantum Liouville equation. For this, we use natural units with $\hbar = c =1$ and stay in the interaction picture throughout the entire article. The interaction picture Liouville equation is given by \cite{Breuer2002}
\begin{eqnarray}
\label{eq:LiouvilleInt}
\frac{\partial}{\partial t} \hat{\rho}(t) &=& -\mathrm{i}[\hat{H}_I(t),\hat{\rho}(t)]~,
\end{eqnarray}
where $\hat{H}_I(t)$ comprises all Hamiltonians that describe self-interactions of the considered systems in the Hilbert space $\mathcal{H}$ and interactions between them. For example, if we consider two systems $A$ and $B$, then $\hat{H}_I(t) = \hat{H}_A(t) + \hat{H}_B(t) + \hat{H}_{AB}(t)$, where $\hat{H}_A(t)$ and $\hat{H}_B(t)$ are the self-interactions of $A$ and $B$, respectively, and $\hat{H}_{AB}(t)$ describes interactions between both systems. Eq.~(\ref{eq:LiouvilleInt}) can simply be solved by either
\begin{eqnarray}
\label{eq:Sol1}
 \hat{\rho}(t) &=& \hat{\rho}(0) -\mathrm{i} \int\limits_0^t d\tau [\hat{H}_I(\tau),\hat{\rho}(\tau)]
\end{eqnarray}
or 
\begin{eqnarray}\label{eq:Sol2}
\hat{\rho}(t) &=& (\mathrm{T} e^{-\mathrm{i}\int^t_0 d\tau \hat{H}_I(\tau)})\hat{\rho}(0)(\tilde{\mathrm{T}} e^{\mathrm{i}\int^t_0 d\tau \hat{H}_I(\tau)})
\end{eqnarray}
with $\mathrm{T}$ and $\tilde{\mathrm{T}}$ standing for time-ordering and anti-time-ordering, respectively. Note that, for notational convenience, we use $t=0$ as the initial time in this article. While Eq.~(\ref{eq:Sol1}) serves as the starting point for deriving the Lindblad master equation \cite{Manzano_2020}, Eq.~(\ref{eq:Sol2}) has the advantage that it does not depend on integrations of the density matrix in all times from $0$ to $t$.

\begin{figure}[htbp]
\begin{center}
\includegraphics[scale=0.5]{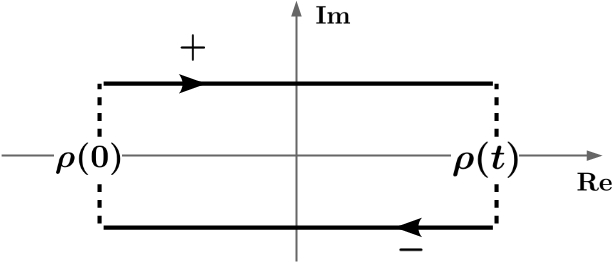}
\caption{Taken from Refs.~\cite{Kading:2022jjl,Kading:2022hhc}; schematic depiction of the closed time path for a density matrix $\rho$ evolving from an initial time $0$ to a final time $t$ and backwards; all degrees of freedom are doubled and associated with either the $+$ or $-$ branch of the closed time path. Formally, the time coordinate on the $\pm$ branch is shifted by $\pm\mathrm{i}\varepsilon$, and operators are time-ordered/anti-time-ordered on the $+$ branch/ $-$ branch.}
\label{fig:CTP}
\end{center}
\end{figure}

Next, we lend tools from TFD \cite{Takahasi:1974zn,Arimitsu:1985ez,Arimitsu:1985xm, Khanna} in order to get a different perspective on Eq.~(\ref{eq:LiouvilleInt}). In our context, we can see TFD as an algebraic version of the Schwinger-Keldysh formalism \cite{Schwinger,Keldysh}, which is formulated on a closed time path; see Fig.~\ref{fig:CTP}. The Schwinger-Keldysh formalism offers a powerful way of computing expectation values of operators on a single time slice in terms of path integrals, and serves as the basis for the famous Feynman-Vernon influence functional \cite{Feynman}.
TFD is set in a doubled Hilbert space $\widehat{\mathcal{H}} := \mathcal{H}^+ \otimes \mathcal{H}^-$, where $\mathcal{H}^+$ and $\mathcal{H}^-$ each are the Hilbert space $\mathcal{H}$ that we have considered earlier, but are now associated to the $+$ and $-$ branch of the closed time path, respectively. In practice, this means that we can define operators 
\begin{eqnarray}
\label{eq:TFDoper}
\hat{\mathcal{O}}^+ &=& \hat{\mathcal{O}} \otimes \hat{\mathbb{I}}~~,~~
\hat{\mathcal{O}}^- = \hat{\mathbb{I}} \otimes \hat{\mathcal{O}}^\mathcal{T}~,
\end{eqnarray}
with $\mathcal{T}$ indicating time-reversal, that can act on states 
\begin{eqnarray}
    \kket{\mathfrak{a}_+,\mathfrak{b}_-} &:=& \ket{\mathfrak{a}} \otimes \ket{\mathfrak{b}}
\end{eqnarray}
in some basis of the doubled Hilbert space. Furthermore, we can define a state \cite{Arimitsu:1985ez}
\begin{eqnarray}
\label{eq:1state}
    \kket{1} &:=& \sumint_\mathfrak{h} \kket{\mathfrak{h}_+,\mathfrak{h}_-}~,
\end{eqnarray}
where $\ket{\mathfrak{h}}$ is a normalized basis state of the Hilbert space $\mathcal{H}$, $\sumint$ is a $\sum$ for a discrete basis and/or a $\int$ for a continuous basis with the inner product $\braket{\mathfrak{h}|\mathfrak{g}} = \delta(\mathfrak{h} - \mathfrak{g})$ with $\delta(\mathfrak{h} - \mathfrak{g})$ either being the Kronecker delta (discrete basis) or the Dirac delta (continuous basis). Note that the state in Eq.~(\ref{eq:1state}) is time- and picture-independent \cite{Kading:2022jjl}. Using Eq.~(\ref{eq:1state}), TFD enables us to compute expectation values of operators as 
\begin{eqnarray}
\label{eq:OpTrace}
\braket{\hat{\mathcal{O}}(t)} &=& \text{Tr}\hat{\mathcal{O}}(t)\hat{\rho}(t) = \bbrakket{1| \hat{\mathcal{O}}^+(t)\hat{\rho}^+(t) |1} = \bbrakket{1| \hat{\mathcal{O}}^-(t)\hat{\rho}^-(t) |1}^\ast~.
\end{eqnarray}
In addition, the quantum Liouville equation (\ref{eq:LiouvilleInt}) can be expressed in a Schr\"odinger-like form
\begin{eqnarray}
\label{eq:SchrLiouvEq}
\frac{\partial}{\partial t} \hat{\rho}^+(t)\kket{1} &=& -\mathrm{i}\widehat{H}_I(t)\hat{\rho}^+(t)\kket{1}~,
\end{eqnarray}
where $\widehat{H}_I(t) := \hat{H}_I(t) \otimes \hat{\mathbb{I}} - \hat{\mathbb{I}} \otimes \hat{H}_I(t)$; see App.~\ref{AppQL}. We can solve Eq.~(\ref{eq:SchrLiouvEq}) either by
\begin{eqnarray}
\label{eq:Sol1a}
 \hat{\rho}^+(t)\kket{1} &=& \hat{\rho}^+(0)\kket{1} -\mathrm{i} \int\limits_0^t d\tau \widehat{H}_I(\tau)\hat{\rho}^+(\tau)\kket{1}
\end{eqnarray}
or by
\begin{eqnarray}
\label{eq:Sol2a}
\hat{\rho}^+(t)\kket{1} &=& \text{T} \exp\left[{-\mathrm{i}\int\limits_{0}^t d\tau\widehat{H}_I(\tau)}\right]\hat{\rho}^+(0)\kket{1}~,
\end{eqnarray}
corresponding to the solutions in Eqs.~(\ref{eq:Sol1}) and (\ref{eq:Sol2}). Since we want to translate these solutions into path integrals when looking at the Fock basis, it is useful to replace the Hamiltonian by introducing an action operator \cite{Kading:2022jjl}
\begin{eqnarray}
\label{eq:DefAction}
\widehat{H}_I(\tau) &=& \widehat{H}(\tau) - \widehat{H}_0(\tau)
\nonumber
\\
&=&
\frac{\partial}{\partial \tau} \widehat{S}_0(\tau) - \frac{\partial}{\partial \tau} \widehat{S}(\tau)
\nonumber
\\
&=&
\frac{\partial}{\partial \tau}\widehat{S}_0(\tau) + \sum_i\sum_{a=\pm} a\,\dot{\phi}_i^a \pi_i^a - \left[\frac{\partial}{\partial \tau} \widehat{S}(\tau) + \sum_i\sum_{a=\pm} a\,\dot{\phi}_i^a \pi_i^a\right]
\nonumber
\\
&=&
\frac{d}{d \tau} \widehat{S}_0(\tau) - \frac{d}{d \tau} \widehat{S}(\tau)
\nonumber
\\
&=&
- \frac{d}{d\tau} \widehat{S}_I(\tau)
~,
\end{eqnarray}
where $\widehat{S}_I(t) := \hat{S}_I(t) \otimes \hat{\mathbb{I}} - \hat{\mathbb{I}} \otimes \hat{S}_I(t)$, and $\widehat{S}_I(t,0)=:\widehat{S}_I(t)$ for notational convenience. Above, the sum over $i$ extends over all dynamical field degrees of freedom, each of which is collectively denoted as $\phi_i^a$. Eqs.~(\ref{eq:Sol1a}) and (\ref{eq:Sol2a}) become
\begin{eqnarray}
\label{eq:Sol1b}
 \hat{\rho}^+(t)\kket{1} &=& \hat{\rho}^+(0)\kket{1} +\mathrm{i} \int\limits_{0}^t d\tau \frac{d}{d\tau}\widehat{S}_I(\tau)\hat{\rho}^+(\tau)\kket{1}
\end{eqnarray}
and
\begin{eqnarray}
\label{eq:Sol2b}
\hat{\rho}^+(t)\kket{1} &=&  \exp\left[{\mathrm{i}\widehat{S}_I(t)}\right]\hat{\rho}^+(0)\kket{1} 
\nonumber
\\
&=&
\sum_{n=0}^\infty \frac{1}{n!}\left[{\mathrm{i}\widehat{S}_I(t)}\right]^n
\hat{\rho}^+(0)\kket{1} 
~,
\end{eqnarray}
where we have used $\text{T} \widehat{S}^n_I(t) = \widehat{S}^n_I(t)$. Similarly to Eqs.~(\ref{eq:Sol1}) and (\ref{eq:Sol2}), the right-hand side of Eq.~(\ref{eq:Sol1b}) depends on integrations of the density matrix in the entire interval $[0,t]$, while the right-hand side of Eq.~(\ref{eq:Sol2b}) does not. Since Eqs.~(\ref{eq:Sol1b}) and (\ref{eq:Sol2b}) are equivalent, we conclude that all the memory of the considered systems after the initial time must be encoded in the sum over all powers of the action operator $\widehat{S}_I(t)$. 

Using a series expansion, we can further illustrate the equivalence between Eqs.~(\ref{eq:Sol1b}) and (\ref{eq:Sol2b}). Iteratively substituting Eq.~(\ref{eq:Sol1b}) into itself, we find
\begin{eqnarray}
\label{eq:Sol1c}
 \hat{\rho}^+(t)\kket{1} &=& \hat{\rho}^+(0)\kket{1}
 + \mathrm{i} \widehat{S}_I(t) \hat{\rho}^+(0)\kket{1}
-  \frac{1}{2}\widehat{S}^2_I(t) \hat{\rho}^+(0)\kket{1}
\nonumber
\\
&&
- \mathrm{i} \int\limits_{0}^t d\tau_1 \frac{d}{d\tau_1}\widehat{S}_I(\tau_1)\int\limits_{0}^{\tau_1} d\tau_2 \frac{d}{d\tau_2}\widehat{S}_I(\tau_2)\int\limits_{0}^{\tau_2} d\tau_3 \frac{d}{d\tau_3}\widehat{S}_I(\tau_3)\hat{\rho}^+(\tau_3)\kket{1}~.
\end{eqnarray}
In addition, from Eq.~(\ref{eq:Sol2b}), we get
\begin{eqnarray}
\label{eq:Sol2c}
\hat{\rho}^+(t)\kket{1} 
&=&
\hat{\rho}^+(0)\kket{1}
 + \mathrm{i} \widehat{S}_I(t) \hat{\rho}^+(0)\kket{1}
-  \frac{1}{2}\widehat{S}^2_I(t) \hat{\rho}^+(0)\kket{1}
+
\sum_{n=3}^\infty \frac{1}{n!}\left[{\mathrm{i}\widehat{S}_I(t)}\right]^n
\hat{\rho}^+(0)\kket{1} 
~.
\end{eqnarray}
As anticipated, up to second order, and clearly up to any higher order we could consider, the terms in Eqs.~(\ref{eq:Sol1c}) and (\ref{eq:Sol2c}) are identical. 

Next, we show that the dynamical map acting on $\hat{\rho}^+(0)\kket{1}$ on the right-hand side of Eq.~(\ref{eq:Sol2b}) is indeed divisible as we would expect due to the cocycle property of unitary evolution operators \cite{Loubenets2020}. For this, we consider an intermediate time $t'$ with $0 < t' < t$, and recall that $\widehat{S}_I(b,a)$ is defined as an integral from time $a$ to time $b$ over a Hamiltonian, such that $\widehat{S}_I(b,a) = \widehat{S}_I(b,c) + \widehat{S}_I(c,a)$. We can now prove divisibility by showing that
\begin{eqnarray}
\label{eq:FirstOp}
\exp\left[{\mathrm{i}\widehat{S}_I(t)}\right]&=&\exp\left[{\mathrm{i}\widehat{S}_I(t,t')}\right]\exp\left[{\mathrm{i}\widehat{S}_I(t')}\right]~.
\end{eqnarray}
For $t'=0$ (or $t'=t$), Eq.~(\ref{eq:FirstOp}) is trivially true. Hence, we have to show that the derivative $\displaystyle\frac{d}{dt'}$ of this equation's right-hand side vanishes. Due to
\begin{eqnarray}
\label{eq:TDTOP}
 \frac{d}{dt'}\exp\left[{\mathrm{i}\widehat{S}_I(t,t')}\right]&=&-\mathrm{i}\exp\left[{\mathrm{i}\widehat{S}_I(t,t')}\right]\frac{d}{dt'} \widehat{S}_I(t') ~,~~~
 \frac{d}{dt'}\exp\left[{\mathrm{i}\widehat{S}_I(t')}\right]\,=\,\mathrm{i}\frac{d}{dt'} \widehat{S}_I(t')\exp\left[{\mathrm{i}\widehat{S}_I(t')}\right]~,~~~~~~
\end{eqnarray}
see App.~\ref{AppA}, we have indeed
\begin{eqnarray}
 \frac{d}{dt'}\left(\exp\left[{\mathrm{i}\widehat{S}_I(t,t')}\right]\exp\left[{\mathrm{i}\widehat{S}_I(t')}\right]\right)&=&0~,
\end{eqnarray}
and hence
\begin{eqnarray}
\hat{\rho}^+(t)\kket{1} &=& \exp\left[{\mathrm{i}\widehat{S}_I(t)}\right]\hat{\rho}^+(0)\kket{1} 
\nonumber
\\
&=&
\exp\left[{\mathrm{i}\widehat{S}_I(t,t')}\right]\exp\left[{\mathrm{i}\widehat{S}_I(t')}\right]\hat{\rho}^+(0)\kket{1} 
~,
\end{eqnarray}
which proves divisibility.


\subsection{General basis}
\label{sec:Closed}

In order to make quantitative predictions, it is often useful to know the density matrix elements in a particular basis. The total density matrix can be expressed as a sum and/or an integral over all basis vectors of the Hilbert space $\mathcal{H}$:
\begin{eqnarray}
\label{eq:TotDensExp}
    \hat{\rho}(t) &=& \sumint_{\mathfrak{g}\mathfrak{h}} \rho(\mathfrak{g};\mathfrak{h};t) \ket{\mathfrak{g};t}\bra{\mathfrak{h};t}~,
\end{eqnarray}
where
$\rho(\mathfrak{g};\mathfrak{h};t) :=
\bra{\mathfrak{g};t} \hat{\rho}(t)\ket{\mathfrak{h};t}$ are the density matrix elements with $\rho^\ast(\mathfrak{h};\mathfrak{g};t) = \rho(\mathfrak{g};\mathfrak{h};t)$. Following Ref.~\cite{Kading:2022hhc}, we can directly compute a matrix element $\rho(\mathfrak{g};\mathfrak{h};t)$ at time $t$ by using Eq.~(\ref{eq:OpTrace}), such that 
\begin{eqnarray}
    \rho(\mathfrak{g};\mathfrak{h};t) = \langle \mathfrak{g};t|\hat{\rho}(t)|\mathfrak{h};t\rangle = \bbra{1} 
    (\ket{\mathfrak{h};t}\bra{\mathfrak{g};t} \otimes \hat{\mathbb{I}})\hat{\rho}^+(t)
    \kket{1}~,
\end{eqnarray}
and subsequently substituting either Eq.~(\ref{eq:Sol2b}) or Eq.~(\ref{eq:Sol2c}), and (\ref{eq:TotDensExp}):
\begin{eqnarray}
\label{eq:GenBssExprExEx}
    \rho(\mathfrak{g};\mathfrak{h};t) &=& \bbra{\mathfrak{g}_+;\mathfrak{h}_-;t} 
  \exp\left[{\mathrm{i}\widehat{S}_I(t)}\right]\sumint_{\mathfrak{a}\mathfrak{b}}  \rho(\mathfrak{a};\mathfrak{b};0) \kket{\mathfrak{a}_+;\mathfrak{b}_-;0} \nonumber\\
 &\approx& \bbra{\mathfrak{g}_+;\mathfrak{h}_-;t} 
  \left[\hat{\mathbb{I}} + \mathrm{i} \widehat{S}_I(t) -  \frac{1}{2}\widehat{S}^2_I(t) \right]\sumint_{\mathfrak{a}\mathfrak{b}} \rho(\mathfrak{a};\mathfrak{b};0) \kket{\mathfrak{a}_+;\mathfrak{b}_-;0}~.
\end{eqnarray}
While in many cases we could simply evaluate Eq.~(\ref{eq:GenBssExprExEx}) in its given form, in others, e.g., in quantum field theory, it can be useful to take some additional steps in order rewrite this equation in terms of path integrals. We will do so in the next subsection.


\subsection{Momentum basis in Fock space}
\label{sec:MomBasisClos}

Next, we make closer contact with the method presented in Ref.~\cite{Kading:2022hhc} and consider a closed system consisting of two different scalar field species $\phi$ and $\chi$. Each of these species has its associated Fock space, such that the Hilbert space $\mathcal{H}$ of the total system is the direct product of the two Fock spaces. We choose to work in a momentum basis, such that the density matrix can be expanded as  
\begin{eqnarray}
\label{eq:DensMatrExTot}
\hat{\rho}(t) &=& \sum\limits_{I,J=0}^\infty \frac{1}{I!J!}\int\left(   \prod\limits_{A = 1}^I d\Pi_{K^A}\right)\left(\prod\limits_{B = 1}^J   d\Pi_{L^B} \right)\rho_{I;J}(K^I|L^J|t) \ket{K^I;t}\bra{L^J;t}~,
\end{eqnarray}
where $I := (i_\phi,i_\chi)$ and $J := (j_\phi,j_\chi)$ are bi-indices with factorials $I! := i_\phi ! i_\chi !$ and $J! := j_\phi ! j_\chi !$; we use the short-hand notation $K^I := \mathbf{k}_\phi^{(1)},...,\mathbf{k}_\phi^{(i_\phi)};\mathbf{k}_\chi^{(1)},...,\mathbf{k}_\chi^{(i_\chi)}$ and $L^J := \mathbf{l}_\phi^{(1)},...,\mathbf{l}_\phi^{(j_\phi)};\mathbf{l}_\chi^{(1)},...,\mathbf{l}_\chi^{(j_\chi)}$ for the 3-momenta; and 
\begin{eqnarray}
\prod\limits_{A = 1}^I d\Pi_{K^A} &:=& \left(\prod\limits_{a = 1}^{i_\phi} d\Pi_{\mathbf{k}_\phi^{(a)}}\right)\left(\prod\limits_{a = 1}^{i_\chi} d\Pi_{\mathbf{k}^{(a)}_\chi}\right)~,
\end{eqnarray}
where 
\begin{eqnarray}
\int d\Pi_{\mathbf{k}_\phi} &:=& \int_{\mathbf{k}_\phi} \frac{1}{2E_{\mathbf{k}_\phi}^\phi}
~~,~~
\int_{\mathbf{k}_\phi} := \int \frac{d^3k_\phi}{(2\pi)^3}
\end{eqnarray}
with $E_{\mathbf{k}_\phi}^\phi$ being the on-shell energy associated with the scalar field $\phi$ and the momentum $\mathbf{k}_\phi$. The subscripts on the density matrix elements on the right-hand side of Eq.~(\ref{eq:DensMatrExTot}) label the occupation numbers in Fock space, i.e., the numbers of $\phi$- and $\chi$-particles. For $\phi$, we introduce annihilation and creation operators \cite{Burrage2018}
\begin{eqnarray}
\label{eq:CreatAnni}
\hat{a}^+_{\mathbf{p}}(t) &=& +\mathrm{i}\int_{\mathbf{x}} e^{-\mathrm{i}\mathbf{p}\cdot\mathbf{x}}\partial_{t,E^\phi_{\mathbf{p}}}\hat{\phi}^+_{t,\mathbf{x}}~~,~~
\hat{a}^{+\dagger}_{\mathbf{p}}(t) = -\mathrm{i}\int_{\mathbf{x}} e^{+\mathrm{i}\mathbf{p}\cdot\mathbf{x}} \partial_{t,E^\phi_{\mathbf{p}}}^*\hat{\phi}^+_{t,\mathbf{x}}~,
\nonumber
\\
\hat{a}^-_{\mathbf{p}}(t) &=& -\mathrm{i}\int_{\mathbf{x}} e^{+\mathrm{i}\mathbf{p}\cdot\mathbf{x}}\partial_{t,E^\phi_{\mathbf{p}}}^*\hat{\phi}^-_{t,\mathbf{x}}~~,~~
\hat{a}^{-\dag}_{\mathbf{p}}(t) = +\mathrm{i}\int_{\mathbf{x}} e^{-\mathrm{i}\mathbf{p}\cdot\mathbf{x}}\partial_{t,E^\phi_{\mathbf{p}}}\hat{\phi}^{-}_{t,\mathbf{x}}~,
\end{eqnarray}
where
\begin{eqnarray}
\int_{x} &:=& \int_{\Omega_t} d^4x
\end{eqnarray}
with $\Omega_t := [0,t]\times\mathbb{R}^3$, and $\partial_{t,E^\phi_{\mathbf{p}}} := \overset{\rightarrow}{\partial}_t - \mathrm{i}E^\phi_{\mathbf{p}}$. Further, we introduce $\hat{b}^\dagger$ and $\hat{b}$ as creators and annihilators for $\chi$ of the same form as $\hat{a}^\dagger$ and $\hat{a}$. Note that we have introduced $\hat{\phi}^\pm_{t,\mathbf{x}}:=\hat{\phi}^\pm(t,\mathbf{x})$ for notational convenience.

We now want to find an expression for the density matrix element $\rho_{G;H}(K^G|L^H|t)$ by using Eq.~(\ref{eq:GenBssExprExEx}) for the chosen Fock basis. This means that we write
\begin{eqnarray}
    \rho_{G;H}(K^G|L^H|t) &=& \bbra{K^G_+;L^H_-;t} 
  \exp\left[{\mathrm{i}\widehat{S}_I(t)}\right]
  \nonumber
  \\
  &&
  \times
  \sum\limits_{I,J=0}^\infty \frac{1}{I!J!}\int\left(   \prod\limits_{A = 1}^I d\Pi_{R^A}\right)\left(\prod\limits_{B = 1}^J   d\Pi_{P^B} \right)\rho_{I;J}(R^I|P^J|0) 
  \kket{R^I_+;P^J_-;0}~.
\end{eqnarray}
Next, we pull out all creation and annihilation operators that, introducing the TFD vacuum vector $\kket{0} := \ket{0} \otimes \ket{0}$, act like 
\begin{eqnarray}\label{eq:Creators}
\hat{a}^{\pm\dagger}_\mathbf{p} \kket{0} =  \kket{\mathbf{p}_\pm}
\,\,\,,\,\,\,\,\,\,\,\,\,
\hat{a}^{\pm}_\mathbf{p} \kket{\mathbf{k}_+;\mathbf{k}_-} =  (2\pi)^3 2 E^\phi_\mathbf{p} \delta(\mathbf{k}-\mathbf{p})\kket{\mathbf{k}_\mp}
\end{eqnarray}
(and similarly for $\hat{b}^\dagger$ and $\hat{b}$). This leaves us with
\begin{eqnarray}
\label{eq:IntermedClossysde}
    \rho_{G;H}(K^G|L^H|t) 
    &=& 
    \sum\limits_{I,J=0}^\infty \frac{1}{I!J!}\int\left(   \prod\limits_{A = 1}^I d\Pi_{R^A}\right)\left(\prod\limits_{B = 1}^J   d\Pi_{P^B} \right)\rho_{I;J}(R^I|P^J|0) 
  \nonumber
  \\
  &&
  \times
  \bbra{0} 
  \hat{a}^+_{K^G}(t) \hat{b}^+_{K^G}(t)\hat{a}^-_{L^H}(t)\hat{b}^-_{L^H}(t)
  \exp\left[{\mathrm{i}\widehat{S}_I(t)}\right]
    \nonumber
  \\
  &&
  ~~~~~~~~~~~~~~~~~~~~~~~~~~~~~~~~~~~~~~~~~~~
  \times
  \hat{a}^{+\dagger}_{R^I}(0) \hat{b}^{+\dagger}_{R^I}(0)\hat{a}^{-\dagger}_{P^J}(0)\hat{b}^{-\dagger}_{P^J}(0)
  \kket{0}~,
\end{eqnarray}
where $\hat{a}^+_{K^G}(t) := \hat{a}^+_{\mathbf{k}_\phi^{(1)}}(t)\,...\,\hat{a}^+_{\mathbf{k}_\phi^{(g_\phi)}}(t)$ and $\hat{b}^+_{K^G}(t) := \hat{b}^+_{\mathbf{k}_\chi^{(1)}}(t)\,...\,\hat{b}^+_{\mathbf{k}_\chi^{(g_\chi)}}(t)$, etc. Substituting Eq.~(\ref{eq:CreatAnni}) into Eq.~(\ref{eq:IntermedClossysde}), we obtain \cite{Kading:2022hhc}:
\begin{eqnarray}
\label{eq:ToDensOperExp}
\rho_{G;H}(K^G|L^H|t)
&=&
\sum\limits_{I,J=0}^\infty \frac{\mathrm{i}^{G+J}(-\mathrm{i})^{H+I}}{I!J!}
\lim_{\substack{X^{0,G},X^{0\prime,H}\,\to\, t^{+}\\Y^{0,I},Y^{0\prime,J}\,\to\, 0^-}}
\int \left(   \prod\limits_{A = 1}^I d\Pi_{R^A}\right)\left(\prod\limits_{B = 1}^J   d\Pi_{P^B} \right) \rho_{I;J}(R^I|P^J|0) 
\nonumber
\\
&&\,\,\,\,\,\,
\times 
\int_{X^G X^{\prime H}Y^I Y^{\prime J}}
\exp\left\{-\mathrm{i}\big(K^G  X^G
- L^H  X^{\prime H}\big)
+ \mathrm{i}\big(R^IY^I - P^JY^{\prime J}\big)\right\}
\nonumber
\\
&&\,\,\,\,\,\,
\times 
\left(   \prod\limits_{A = 1}^G
\partial_{X^{0,A},E_{K^{A}}} \right)
\left(  \prod\limits_{B = 1}^H
\partial_{X^{0\prime,B},E_{L^B}}^*\right)
\left(\prod\limits_{C = 1}^I \partial_{Y^{0,C},E_{R^C}}^* \right)
\left(\prod\limits_{D = 1}^J 
\partial_{Y^{0\prime,D},E_{P^D}} 
\right)
\nonumber
\\
&&\,\,\,\,\,\,
\times 
\bbra{0} 
    \hat{\phi}^+_{X^G}\hat{\chi}^+_{X^G}\hat{\phi}^-_{X^{\prime H}}\hat{\chi}^-_{X^{\prime H}}
  \exp\left[{\mathrm{i}\widehat{S}_I(t)}\right]
  \hat{\phi}^+_{Y^I}\hat{\chi}^+_{Y^I}\hat{\phi}^-_{Y^{\prime J}}\hat{\chi}^-_{Y^{\prime J}}
  \kket{0}
~,~
\end{eqnarray}
where we have defined $\mathrm{i}^G := \mathrm{i}^{g_\phi + g_\chi}$ for the imaginary unit; $X^{0,G} := x^0_{\phi,(1)},...,x^0_{\phi,(g_\phi)}; x^0_{\chi,(1)},...,x^0_{\chi,(g_\chi)}$ for the time coordinates; $X^G := \mathbf{x}_{\phi,(1)},...,\mathbf{x}_{\phi,(g_\phi)};\mathbf{x}_{\chi,(1)},...,\mathbf{x}_{\chi,(g_\chi)}$ for the spatial coordinates; and $\hat{\phi}^+_{X^G} := \hat{\phi}^+_{x_{\phi,(1)}} ...\, \hat{\phi}^+_{x_{\phi,(g_\phi)}}$ and $\hat{\chi}^+_{X^G} := \hat{\chi}^+_{x_{\chi,(1)}} ...\, \hat{\chi}^+_{x_{\chi,(g_\chi)}}$ etc. for the field operators. In addition, we have defined
\begin{eqnarray}
\prod\limits_{A = 1}^G
\partial_{X^{0,A},E_{K^{A}}} &:=& \partial_{x^{0}_{\phi,(1)},E^\phi_{\mathbf{k}_\phi^{(1)}}} ...\, \partial_{x^{0}_{\phi,(g_\phi)},E^\phi_{\mathbf{k}_\phi^{(g_\phi)}}}
\partial_{x^{0}_{\chi,(1)},E^\chi_{\mathbf{k}_\chi^{(1)}}} ...\, \partial_{x^{0}_{\chi,(g_\chi)},E^\chi_{\mathbf{k}_\chi^{(g_\chi)}}}
~,
\end{eqnarray}
and introduced limits for the time coordinates in order to assure that the Klein-Gordon operators act only on the fields external to the actions. Next, we write $\widehat{S}_I(t) = \widehat{S}_\phi(t) + \widehat{S}_\chi(t) + \widehat{S}_{\phi\chi}(t)$, where $\widehat{S}_\phi(t)$ and $\widehat{S}_\chi(t)$ describe the self-interactions of $\phi$ and $\chi$, respectively, and $\widehat{S}_{\phi\chi}(t)$ represents the interaction between both systems. Since the last line of Eq.~(\ref{eq:ToDensOperExp}) is a correlation function, we can express it in terms of path integrals, such that we obtain
\begin{eqnarray}
\label{eq:CloPath}
\rho_{G;H}(K^G|L^H|t)
&=&
\sum\limits_{I,J=0}^\infty \frac{\mathrm{i}^{G+J}(-\mathrm{i})^{H+I}}{I!J!}
\lim_{\substack{X^{0,G},X^{0\prime,H}\,\to\, t^{+}\\Y^{0,I},Y^{0\prime,J}\,\to\, 0^-}}
\int \left(   \prod\limits_{A = 1}^I d\Pi_{R^A}\right)\left(\prod\limits_{B = 1}^J   d\Pi_{P^B} \right) \rho_{I;J}(R^I|P^J|0) 
\nonumber
\\
&&\,\,\,\,\,\,
\times 
\int_{X^G X^{\prime H}Y^I Y^{\prime J}}
\exp\left\{-\mathrm{i}\big(K^G  X^G
- L^H  X^{\prime H}\big)
+ \mathrm{i}\big(R^IY^I - P^JY^{\prime J}\big)\right\}
\nonumber
\\
&&\,\,\,\,\,\,
\times 
\left(   \prod\limits_{A = 1}^G
\partial_{X^{0,A},E_{K^{A}}} \right)
\left(  \prod\limits_{B = 1}^H
\partial_{X^{0\prime,B},E_{L^B}}^*\right)
\left(\prod\limits_{C = 1}^I \partial_{Y^{0,C},E_{R^C}}^* \right)
\left(\prod\limits_{D = 1}^J 
\partial_{Y^{0\prime,D},E_{P^D}} 
\right)
\nonumber
\\
&&\,\,\,\,\,\,
\times 
\int\mathcal{D}\phi^{\pm} \mathcal{D}\chi^{\pm}
e^{\mathrm{i}\widehat{S}[\phi] + \mathrm{i}\widehat{S}[\chi]}
   \phi^+_{X^G}\chi^+_{X^G}\phi^-_{X^{\prime H}}\chi^-_{X^{\prime H}}
  e^{\mathrm{i}\widehat{S}_\phi(t) + \mathrm{i}\widehat{S}_\chi(t) + \mathrm{i}\widehat{S}_{\phi\chi}(t)}
  \phi^+_{Y^I}\chi^+_{Y^I}\phi^-_{Y^{\prime J}}\chi^-_{Y^{\prime J}} \nonumber\\
  &\approx&
\sum\limits_{I,J=0}^\infty \frac{\mathrm{i}^{G+J}(-\mathrm{i})^{H+I}}{I!J!}
\lim_{\substack{X^{0,G},X^{0\prime,H}\,\to\, t^{+}\\Y^{0,I},Y^{0\prime,J}\,\to\, 0^-}}
\int \left(   \prod\limits_{A = 1}^I d\Pi_{R^A}\right)\left(\prod\limits_{B = 1}^J   d\Pi_{P^B} \right) \rho_{I;J}(R^I|P^J|0) 
\nonumber
\\
&&\,\,\,\,\,\,
\times 
\int_{X^G X^{\prime H}Y^I Y^{\prime J}}
\exp\left\{-\mathrm{i}\big(K^G  X^G
- L^H  X^{\prime H}\big)
+ \mathrm{i}\big(R^IY^I - P^JY^{\prime J}\big)\right\}
\nonumber
\\
&&\,\,\,\,\,\,
\times 
\left(   \prod\limits_{A = 1}^G
\partial_{X^{0,A},E_{K^{A}}} \right)
\left(  \prod\limits_{B = 1}^H
\partial_{X^{0\prime,B},E_{L^B}}^*\right)
\left(\prod\limits_{C = 1}^I \partial_{Y^{0,C},E_{R^C}}^* \right)
\left(\prod\limits_{D = 1}^J 
\partial_{Y^{0\prime,D},E_{P^D}} 
\right)
\nonumber
\\
&&\,\,\,\,\,\,
\times 
\int\mathcal{D}\phi^{\pm} \mathcal{D}\chi^{\pm}
e^{\mathrm{i}\widehat{S}[\phi] + \mathrm{i}\widehat{S}[\chi]}
   \phi^+_{X^G}\chi^+_{X^G}\phi^-_{X^{\prime H}}\chi^-_{X^{\prime H}}
  \bigg\{1 + \mathrm{i} \widehat{S}_\phi[\phi;t] + \mathrm{i} \widehat{S}_\chi[\chi;t] 
  \nonumber
\\
&&\,\,\,\,\,\,\,\,\,\,\,\,
  + \mathrm{i} \widehat{S}_{\phi\chi}[\phi,\chi;t] 
  -  \frac{1}{2}\bigg[
\widehat{S}^2_\phi[\phi;t] + \widehat{S}^2_\chi[\chi;t] + \widehat{S}^2_{\phi\chi}[\phi,\chi;t]
+ 2 \widehat{S}_\phi[\phi;t] \widehat{S}_\chi[\chi;t]
 \nonumber
\\
&&\,\,\,\,\,\,\,\,\,\,\,\,\,\,\,\,\,\,\,\,\,\,\,\,\,\,\,
+ 2\widehat{S}_\phi[\phi;t]\widehat{S}_{\phi\chi}[\phi,\chi;t]
 + 2\widehat{S}_\chi[\chi;t]\widehat{S}_{\phi\chi}[\phi,\chi;t] 
 \bigg]
  \bigg\}
  \phi^+_{Y^I}\chi^+_{Y^I}\phi^-_{Y^{\prime J}}\chi^-_{Y^{\prime J}}
~,
\end{eqnarray}
where $\mathcal{D}\phi^{\pm} := \mathcal{D}\phi^{+} \mathcal{D}\phi^{-}$; $\widehat{S}[\phi] := S[\phi^+] - S[\phi^-]$ is the free action functional of $\phi$ with $S[\phi^+] = \int_{\Omega_t} \mathcal{L}[\phi^+]$ for the Lagrangian $\mathcal{L}[\phi^+]$; $\phi^+_{X^G} := \phi^+_{x_{\phi,(1)}} ...\, \phi^+_{x_{\phi,(g_\phi)}}$; and analogous definitions hold for $\chi$. $\widehat{S}_\phi[\phi;t]$, $\widehat{S}_\chi[\chi;t]$ and $\widehat{S}_{\phi\chi}[\phi,\chi;t]$ are the self-interaction action functionals for $\phi$ and $\chi$, and the interaction action functional. They are defined in the same manner as the free action functionals. 
The last three lines in Eq.~(\ref{eq:CloPath}) can be evaluated by using Wick's theorem \cite{Wick}, which gives rise to products of different types of two-point functions. Such expressions can then be depicted in terms of Feynman diagrams. However, it should be noted that, at least for closed systems and as long as we consider only vacuum expectation values for zero temperature, only contractions of two $+$-labelled or two $-$-labelled fields are valid, and mixed contractions vanish \cite{Kading:2022jjl}.

\begin{figure}[htbp]
\centering
\subfloat[][]{\includegraphics[scale=0.40]{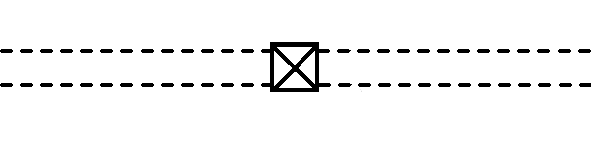}}
\qquad
\subfloat[][]{\includegraphics[scale=0.40]{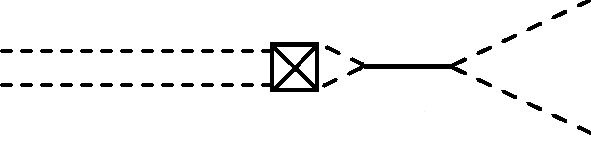}}

\subfloat[][]{\includegraphics[scale=0.40]{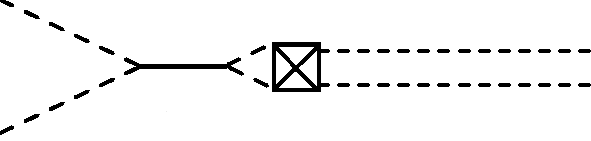}}
\qquad
\subfloat[][]{\includegraphics[scale=0.40]{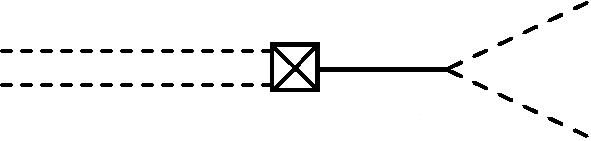}}

\subfloat[][]{\includegraphics[scale=0.40]{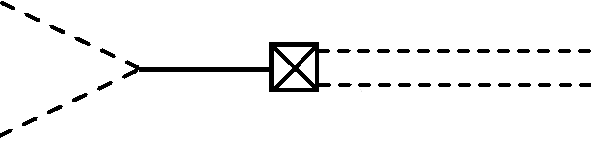}}
\qquad
\subfloat[][]{\includegraphics[scale=0.40]{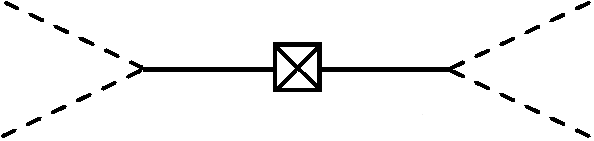}}

\caption{\label{fig:diagrams1} (f) was taken from Ref.~\cite{Kading:2022hhc}, and (a)-(e) are based on it. (a)-(f) show all possible, non-divergent diagrams up to second order in $\lambda$ that contribute to density matrix elements of the form $\rho_{0,2;0,2}(t)$ for the interaction $\lambda\phi\chi^2$. The crossed box represents an insertion of the initial density matrix element, a solid line is a $\phi$ propagator, and a dashed line is a $\chi$ propagator. (a)-(c) arise from the initial condition $\rho_{0,2;0,2}(0)$, (d) from $\rho_{0,2;1,0}(0)$, (e) from $\rho_{1,0;0,2}(0)$, and (f) from $\rho_{1,0;1,0}(0)$.
}
\end{figure}

\begin{figure}[htbp]
\centering
\subfloat[][]{\includegraphics[scale=0.40]{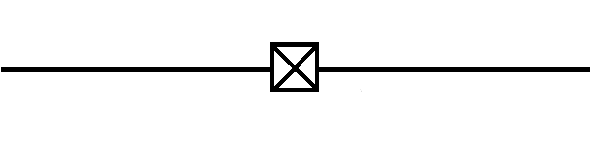}}
\qquad
\subfloat[][]{\includegraphics[scale=0.40]{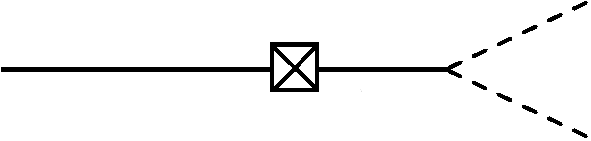}}

\subfloat[][]{\includegraphics[scale=0.40]{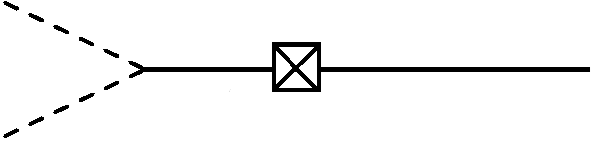}}
\qquad
\subfloat[][]{\includegraphics[scale=0.40]{1to2alternativ.png}}

\subfloat[][]{\includegraphics[scale=0.40]{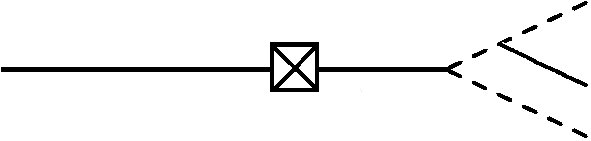}}
\qquad
\subfloat[][]{\includegraphics[scale=0.40]{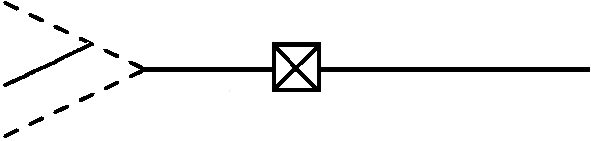}}

\caption{\label{fig:diagrams} (b) and (d) were taken from Ref.~\cite{Kading:2022hhc}, and (a), (c), (e) and (f) are based on these figures. (a)-(f) show all possible, non-divergent diagrams that can be obtained at up to second order in $\lambda$ for the interaction $\lambda\phi\chi^2$ and the only non-vanishing initial density matrix element being $\rho_{1,0;1,0}(0)$. (a) corresponds to 
$\rho_{1,0;1,0}(t)$, (b) to $\rho_{1,0;0,2}(t)$, (c) to $\rho_{0,2;1,0}(t)$, (d) to $\rho_{0,2;0,2}(t)$, (e) to $\rho_{1,0;1,2}(t)$, and (f) to $\rho_{1,2;1,0}(t)$.
}
\end{figure}

Eq.~(\ref{eq:CloPath}) offers us a direct way of computing total density matrix elements in Fock space. At first glance, this equation seems to be difficult to apply if all or many of the Fock space states are occupied at the initial time. However, due to our restriction to second order in perturbation theory, there are many instances when Eq.~(\ref{eq:CloPath}) is powerful even in such cases. For example, if we consider an interaction $\lambda\phi\chi^2$ with coupling constant $\lambda$ and want to compute the density matrix elements $\rho_{0,2;0,2}(t)$, which correspond to nil $\phi$-particles and two $\chi$-particles at the final time, then only initial density matrix elements of the forms $\rho_{0,2;0,2}(0)$, $\rho_{0,2;1,0}(0)$, $\rho_{1,0;0,2}(0)$, and $\rho_{1,0;1,0}(0)$ need to be considered at second order in $\lambda$; see Fig.~\ref{fig:diagrams1}.
Vice versa, if only initial density matrix elements $\rho_{1,0;1,0}(0)$ are non-vanishing, i.e.,~those that correspond to a single $\phi$-particle without any correlations to other states in $\mathcal{H}$, then up to second order in $\lambda$, only density matrix elements of the form $\rho_{1,0;1,0}(t)$, $\rho_{1,0;0,2}(t)$, $\rho_{0,2;1,0}(t)$, $\rho_{0,2;0,2}(t)$, $\rho_{1,0;1,2}(t)$, and $\rho_{1,2;1,0}(t)$ are relevant for us at the final time; see Fig.~\ref{fig:diagrams}. 
In a similar manner, this logic also holds for reduced density matrices of open quantum systems, which is why we do not discuss this issue again in Sec.~\ref{sec:Open}.


\section{Open systems}
\label{sec:Open}

Now we consider open quantum systems. This means that the total Hilbert space is separated into two Hilbert spaces, one for the system $\mathcal{S}$ and one for the environment $\mathcal{E}$, such that $\mathcal{H} = \mathcal{S} \otimes \mathcal{E}$. In order to find the reduced density matrix for the system, we make the usual assumption that there are no initial correlations between system and environment, which means that $\hat{\rho}(0) = \hat{\rho}_\mathcal{S}(0) \otimes \hat{\rho}_\mathcal{E}(0)$, and then trace out the environmental degrees of freedom, i.e., $\hat{\rho}_{\mathcal{S}}(t) = \mathrm{Tr}_\mathcal{E}  \hat{\rho}(t)$. From Eq.~(\ref{eq:OpTrace}) we already know how to take traces in TFD. Introducing $\kket{1} = \kket{1}_\mathcal{S}\otimes\kket{1}_\mathcal{E}$, and simply using ${}_\mathcal{E}\bbra{1}$ on Eq.~(\ref{eq:Sol2b}), we find
\begin{eqnarray}
\label{eq:OpSysTraOutM}
\hat{\rho}_\mathcal{S}^+(t)\kket{1}_\mathcal{S}
&=&
{}_\mathcal{E}\bbra{1}\exp\left[\mathrm{i}\widehat{S}_I(t)\right]\hat{\rho}_\mathcal{E}^+(0)\kket{1}_\mathcal{E}\hat{\rho}_\mathcal{S}^+(0)\kket{1}_\mathcal{S}
~.
\end{eqnarray}
We now separate the action operator into self-interaction terms for system and environment, and a term that describes the interaction between both subsystems, such that $\widehat{S}_I(t) = \widehat{S}_\mathcal{S}(t) +  \widehat{S}_\mathcal{E}(t) + \widehat{S}_{\mathcal{SE}}(t)$. Clearly, only the last two of those terms are affected by the trace, and since $\mathrm{Tr}_\mathcal{E}  \hat{\rho}_\mathcal{E}(0) =1$, Eq.~(\ref{eq:OpSysTraOutM}) becomes
\begin{eqnarray}
\label{eq:Sol2cReduMP}
\hat{\rho}_\mathcal{S}^+(t)\kket{1}_\mathcal{S}
&=&\exp\left[\mathrm{i}\widehat{S}_\mathcal{S}(t) + \mathrm{i}\widehat{S}_{IF}(t)\right]\hat{\rho}_\mathcal{S}^+(0)\kket{1}_\mathcal{S}
~,
\end{eqnarray}
where we have introduced the influence action operator $\widehat{S}_{IF}(t)$ via
\begin{eqnarray}
\label{eq:IF}
    \exp\left[\mathrm{i}\widehat{S}_\mathcal{S}(t) + \mathrm{i}\widehat{S}_{IF}(t)\right]&=&{}_\mathcal{E}\bbra{1}\exp\left[\mathrm{i}\widehat{S}_\mathcal{S}(t) + \mathrm{i}\widehat{S}_\mathcal{E}(t) + \mathrm{i}\widehat{S}_{\mathcal{SE}}(t)\right]\hat{\rho}_\mathcal{E}^+(0)\kket{1}_\mathcal{E}
~,
\end{eqnarray}
which for $[\widehat{S}_\mathcal{S}(t),\widehat{S}_{\mathcal{SE}}(t)]=0$ reduces to
\begin{eqnarray}
\label{eq:IFR}
    \exp\left[\mathrm{i}\widehat{S}_{IF}(t)\right]&=&{}_\mathcal{E}\bbra{1}\exp\left[\mathrm{i}\widehat{S}_\mathcal{E}(t) + \mathrm{i}\widehat{S}_{\mathcal{SE}}(t)\right]\hat{\rho}_\mathcal{E}^+(0)\kket{1}_\mathcal{E}
~,
\end{eqnarray}
and describes the impact of the environment on the system. After introducing the shorthand notation 
\begin{eqnarray}
    \braket{\hat{O}} &:=& {}_\mathcal{E}\bbra{1}
\hat{O}\hat{\rho}_\mathcal{E}^+(0)
 \kket{1}_\mathcal{E}~,
\end{eqnarray}
up to second order in perturbation theory, and for $[\widehat{S}_\mathcal{E}(t),\widehat{S}_{\mathcal{SE}}(t)]=0$, the influence action operator takes on the form
\begin{eqnarray}
\label{eq:IFPerturbed}
    \widehat{S}_{IF}(t) &=& -\mathrm{i}\ln\braket{\exp[\mathrm{i}\widehat{S}_\mathcal{E}(t) + \mathrm{i}\widehat{S}_{\mathcal{SE}}(t)]} \nonumber\\
    &\approx& 
\braket{\widehat{S}_\mathcal{E}(t)} + \braket{\widehat{S}_{\mathcal{SE}}(t)} +\frac{\mathrm{i}}{2}
[\braket{\widehat{S}^2_\mathcal{E}(t)} + 2\braket{\widehat{S}_\mathcal{E}(t)\widehat{S}_{\mathcal{SE}}(t)} + \braket{\widehat{S}^2_{I;\mathcal{SE}}(t)}
]
\nonumber
\\
&&
- \frac{\mathrm{i}}{2}
\left[
\braket{\widehat{S}_\mathcal{E}(t)}^2 
+2 \braket{\widehat{S}_\mathcal{E}(t)}\braket{\widehat{S}_{\mathcal{SE}}(t)}
+
\braket{\widehat{S}_{\mathcal{SE}}(t)}^2
\right]~.
\end{eqnarray}
After expanding Eq.~(\ref{eq:IF}), we can write Eq.~(\ref{eq:Sol2cReduMP}) as
\begin{eqnarray}
\label{eq:OQSPerturb}
    \hat{\rho}_\mathcal{S}^+(t)\kket{1}_\mathcal{S}
&\approx&
\bigg\{ 
\hat{\mathbb{I}}+
\mathrm{i} \widehat{S}_\mathcal{S}(t)
- \frac{1}{2}\widehat{S}^2_\mathcal{S}(t) + \mathrm{i}\widehat{S}_{IF}(t) 
-
\frac{1}{2}\widehat{S}_{IF}^2(t)-
\widehat{S}_\mathcal{S}(t)\widehat{S}_{IF}(t) \bigg\}\,\hat{\rho}_\mathcal{S}^+(0)\kket{1}_\mathcal{S} \nonumber\\
&\approx&
\bigg\{ 
\hat{\mathbb{I}}+
\mathrm{i} \widehat{S}_\mathcal{S}(t)
+ \mathrm{i}\braket{\widehat{S}_\mathcal{E}(t)}
+\mathrm{i}\braket{\widehat{S}_\mathcal{SE}(t)} 
-\frac{1}{2}\widehat{S}_\mathcal{S}^2(t)
-\frac{1}{2}\braket{\widehat{S}_\mathcal{E}^2(t)}
-\frac{1}{2}\braket{\widehat{S}_\mathcal{SE}^2(t)} \nonumber
\\
&&-\widehat{S}_\mathcal{S}(t)\braket{\widehat{S}_\mathcal{E}(t)}
-\widehat{S}_\mathcal{S}(t)\braket{\widehat{S}_\mathcal{SE}(t)}
-\braket{\widehat{S}_\mathcal{E}(t)\widehat{S}_\mathcal{SE}(t)}
\bigg\}\,\hat{\rho}_\mathcal{S}^+(0)\kket{1}_\mathcal{S}~.
\end{eqnarray}
This means that we can write the reduced density matrix elements in a general basis as
\begin{eqnarray}
\label{eq:GenBssExprExExExOS}
    \rho_\mathcal{S}(\mathfrak{g};\mathfrak{h};t) &=& {}_\mathcal{S}\bbra{\mathfrak{g}_+;\mathfrak{h}_-;t} 
  \exp\left[\mathrm{i}\widehat{S}_\mathcal{S}(t) + \mathrm{i}\widehat{S}_{IF}(t)\right]\sumint_{\mathfrak{a}\mathfrak{b}} \rho_\mathcal{S}(\mathfrak{a};\mathfrak{b};0) \kket{\mathfrak{a}_+;\mathfrak{b}_-;0}_\mathcal{S}
\end{eqnarray}
or
\begin{eqnarray}
\label{eq:GenBssExprExExOS}
    \rho_\mathcal{S}(\mathfrak{g};\mathfrak{h};t) &\approx& {}_\mathcal{S}\bbra{\mathfrak{g}_+;\mathfrak{h}_-;t} 
\bigg\{ 
\hat{\mathbb{I}}+
\mathrm{i} \widehat{S}_\mathcal{S}(t)
+ \mathrm{i}\braket{\widehat{S}_\mathcal{E}(t)}
+\mathrm{i}\braket{\widehat{S}_\mathcal{SE}(t)} 
-\frac{1}{2}\widehat{S}_\mathcal{S}^2(t)
-\frac{1}{2}\braket{\widehat{S}_\mathcal{E}^2(t)}
-\frac{1}{2}\braket{\widehat{S}_\mathcal{SE}^2(t)} \nonumber
\\
&&-\widehat{S}_\mathcal{S}(t)\braket{\widehat{S}_\mathcal{E}(t)}
-\widehat{S}_\mathcal{S}(t)\braket{\widehat{S}_\mathcal{SE}(t)}
-\braket{\widehat{S}_\mathcal{E}(t)\widehat{S}_\mathcal{SE}(t)}
\bigg\}\sumint_{\mathfrak{a}\mathfrak{b}} \rho_\mathcal{S}(\mathfrak{a};\mathfrak{b};0) \kket{\mathfrak{a}_+;\mathfrak{b}_-;0}_\mathcal{S}~.
\end{eqnarray}
Next, we can check whether the dynamical map on the right-hand side of Eq.~(\ref{eq:Sol2cReduMP}) is divisible. For this, we apply the same logic as for the closed systems in Sec.~\ref{sec:closedsys}. This means that we again consider an intermediate time $t'$ with $0 < t' < t$ and check whether 
\begin{eqnarray}
    \exp\left[\mathrm{i}\widehat{S}_\mathcal{S}(t) + \mathrm{i}\widehat{S}_{IF}(t)\right] &=& 
    \exp\left[\mathrm{i}\widehat{S}_\mathcal{S}(t,t') + \mathrm{i}\widehat{S}_{IF}(t,t')\right]
    \exp\left[\mathrm{i}\widehat{S}_\mathcal{S}(t') + \mathrm{i}\widehat{S}_{IF}(t')\right],
\end{eqnarray}
respectively, for $[\widehat{S}_\mathcal{S}(t),\widehat{S}_{\mathcal{SE}}(t)]=[\widehat{S}_\mathcal{E}(t),\widehat{S}_{\mathcal{SE}}(t)]=0$,
\begin{eqnarray}
\label{eq:ChDivOSEx}
    \exp\left[\mathrm{i}\widehat{S}_{IF}(t)\right] &=& 
    \exp\left[\mathrm{i}\widehat{S}_{IF}(t,t')\right]
    \exp\left[\mathrm{i}\widehat{S}_{IF}(t')\right],
\end{eqnarray}
is true. Again, this equation is clearly fulfilled for $t' =0$ or $t'=t$. Now we take the derivative with respect to $t'$ of the right-hand side of Eq.~(\ref{eq:ChDivOSEx}) while taking into account the definition of the influence action operator in Eq.~(\ref{eq:IFR}). We then find 
\begin{eqnarray}
    &&\hspace{-10mm}\frac{d}{dt'} 
    \bigg\{
\exp\left[\mathrm{i}\widehat{S}_{IF}(t,t')\right]
    \exp\left[\mathrm{i}\widehat{S}_{IF}(t')\right]
    \bigg\} 
\nonumber
\\
&=&
-\mathrm{i}\braket{\exp\left[\mathrm{i}\widehat{S}_\mathcal{E}(t,t') + \mathrm{i}\widehat{S}_{\mathcal{SE}}(t,t')\right]\frac{d}{dt'} 
\left[ 
\widehat{S}_\mathcal{E}(t') + \widehat{S}_{\mathcal{SE}}(t')
\right]}
\braket{\exp\left[\mathrm{i}\widehat{S}_\mathcal{E}(t') + \mathrm{i}\widehat{S}_{\mathcal{SE}}(t')\right]}
\nonumber
\\
&&
+\mathrm{i}\braket{\exp\left[\mathrm{i}\widehat{S}_\mathcal{E}(t,t') + \mathrm{i}\widehat{S}_{\mathcal{SE}}(t,t')\right]}
\braket{
\frac{d}{dt'} 
\left[\widehat{S}_\mathcal{E}(t') + \widehat{S}_{\mathcal{SE}}(t')
\right]
\exp\left[\mathrm{i}\widehat{S}_\mathcal{E}(t') + \mathrm{i}\widehat{S}_{\mathcal{SE}}(t')\right]}~,
\end{eqnarray}
which generally does not vanish. This means that Eq.~(\ref{eq:ChDivOSEx}) is not fulfilled for all values of $t'$, which in turn implies that the dynamical map on the right-hand side of Eq.~(\ref{eq:Sol2cReduMP}) is not divisible. Markovianity implies divisibility \cite{Milz2019}. Since divisibility in Eq.~(\ref{eq:Sol2cReduMP}) is broken, this equation describes non-Markovian dynamics. Eq.~(\ref{eq:Sol2cReduMP}) was derived  from the quantum Liouville equation (\ref{eq:LiouvilleInt}), without any additional assumptions or specifications of a particular system, by tracing out the environmental degrees of freedom. Consequently, it is the most general description of an open quantum system. This means that, without further assumptions or approximations, every open quantum system is non-Markovian. Since the method in Ref.~\cite{Kading:2022jjl} is derived from projecting Eq.~(\ref{eq:Sol2cReduMP}) into a Fock basis, we can now confirm the claim that it is describing non-Markovian dynamics.

As in Sec.~\ref{sec:closedsys}, we can also look at the perturbative case. For open systems, this means that we have to investigate the right-hand side of Eq.~(\ref{eq:OQSPerturb}). Of course, we again find that divisibility is broken as in the exact case in Eq.~(\ref{eq:Sol2cReduMP}). However, the perturbative case gives rise to an interesting additional detail. After applying the Born approximation \cite{Breuer2002} $\hat{\rho}(t) \approx \hat{\rho}_\mathcal{S}(t) \otimes \hat{\rho}_\mathcal{E}$, we find to second order
\begin{eqnarray}
\label{eq:OPSDivbreakI}
    \hat{\rho}_\mathcal{S}^+(t)\kket{1}_\mathcal{S}
&=& \exp\left[\mathrm{i}\widehat{S}_\mathcal{S}(t)\right]
\braket{\exp\left[\mathrm{i}\widehat{S}_\mathcal{E}(t) + \mathrm{i}\widehat{S}_{\mathcal{SE}}(t)\right]}\hat{\rho}_\mathcal{S}^+(0)\kket{1}_\mathcal{S}
\nonumber
\\
&=& \exp\left[\mathrm{i}\widehat{S}_\mathcal{S}(t,t')\right]\exp\left[\mathrm{i}\widehat{S}_\mathcal{S}(t')\right]
\braket{\exp\left[\mathrm{i}\widehat{S}_\mathcal{E}(t,t') + \mathrm{i}\widehat{S}_{\mathcal{SE}}(t,t') + \mathrm{i}\widehat{S}_\mathcal{E}(t') + \mathrm{i}\widehat{S}_{\mathcal{SE}}(t')\right]}\hat{\rho}_\mathcal{S}^+(0)\kket{1}_\mathcal{S}
\nonumber
\\
&\approx& \exp\left[\mathrm{i}\widehat{S}_\mathcal{S}(t,t')\right]\exp\left[\mathrm{i}\widehat{S}_\mathcal{S}(t')\right]
\braket{\exp\left[\mathrm{i}\widehat{S}_\mathcal{E}(t,t') + \mathrm{i}\widehat{S}_{\mathcal{SE}}(t,t')\right]} \nonumber\\
&&\quad
\times\Big(\braket{\exp\left[\mathrm{i}\widehat{S}_\mathcal{E}(t') + \mathrm{i}\widehat{S}_{\mathcal{SE}}(t')\right]} + \braket{\widehat{S}_\mathcal{E}(t,t') + \widehat{S}_{\mathcal{SE}}(t,t')}\braket{\widehat{S}_\mathcal{E}(t') + \widehat{S}_{\mathcal{SE}}(t')} \nonumber\\
&&\qquad- \braket{\left(\widehat{S}_\mathcal{E}(t,t') + \widehat{S}_{\mathcal{SE}}(t,t')\right)\left(\widehat{S}_\mathcal{E}(t') + \widehat{S}_{\mathcal{SE}}(t')\right)}\Big)\,\hat{\rho}_\mathcal{S}^+(0)\kket{1}_\mathcal{S}
\nonumber
\\
&\approx& \exp\left[\mathrm{i}\widehat{S}_\mathcal{S}(t,t')\right]
\braket{\exp\left[\mathrm{i}\widehat{S}_\mathcal{E}(t,t') + \mathrm{i}\widehat{S}_{\mathcal{SE}}(t,t')\right]}\,\hat{\rho}_\mathcal{S}^+(t')\kket{1}_\mathcal{S}
\nonumber
\\
&&+\Big(\braket{\widehat{S}_\mathcal{E}(t,t') + \widehat{S}_{\mathcal{SE}}(t,t')}\braket{\widehat{S}_\mathcal{E}(t') + \widehat{S}_{\mathcal{SE}}(t')} \nonumber\\
&&\qquad- \braket{\left(\widehat{S}_\mathcal{E}(t,t') + \widehat{S}_{\mathcal{SE}}(t,t')\right)\left(\widehat{S}_\mathcal{E}(t') + \widehat{S}_{\mathcal{SE}}(t')\right)}\Big)\,\hat{\rho}_\mathcal{S}^+(0)\kket{1}_\mathcal{S}
~.
\end{eqnarray}
This means that, at second order in perturbation theory and when the Born approximation can be applied, divisibility (and, therefore, Markovianity) is broken by additional terms
\begin{eqnarray}
    \mathfrak{A}(t,t') &:=& 
    \Big(\braket{\left(\widehat{S}_\mathcal{E}(t,t') + \widehat{S}_{\mathcal{SE}}(t,t')\right)\left(\widehat{S}_\mathcal{E}(t') + \widehat{S}_{\mathcal{SE}}(t')\right)} \nonumber\\
&&\qquad- \braket{\widehat{S}_\mathcal{E}(t,t') + \widehat{S}_{\mathcal{SE}}(t,t')}\braket{\widehat{S}_\mathcal{E}(t') + \widehat{S}_{\mathcal{SE}}(t')}\Big)\,\hat{\rho}_\mathcal{S}^+(0)\kket{1}_\mathcal{S}
\end{eqnarray}
that appear in the last two lines of Eq.~(\ref{eq:OPSDivbreakI}). These terms are connected correlation functions of the environment and the system-environment interaction. Generally, these terms will not vanish, which implies that the dynamical map on the right-hand side of Eq.~(\ref{eq:OQSPerturb}) is indeed not divisible. 


\subsection{Momentum basis in Fock space}
\label{sec:MomFockOpen}

Next, we again consider a momentum basis in Fock space. This time, $\phi$ corresponds to the system $\mathcal{S}$ and $\chi$ to the traced-out environment $\mathcal{E}$. At first, we expand the reduced density matrix:
\begin{eqnarray}
\hat{\rho}_\phi(t) &=&  \sum\limits_{i,j=0}^\infty \frac{1}{i!j!} \int\left(   \prod\limits_{a = 1}^id\Pi_{\mathbf{k}^{(a)}}\right)\left(\prod\limits_{b = 1}^j   d\Pi_{\mathbf{l}^{(b)}} \right)\rho_{i;j}(\mathbf{k}^{(1)},...,\mathbf{k}^{(i)};\mathbf{l}^{(1)},...,\mathbf{l}^{(j)};t) 
\nonumber
\\
&&
~~~~~~~~~~~~~~~~~~~~~~~~~~~~~~~~~~~~~~~~~~~~~~~~~~~
\times
\ket{\mathbf{k}^{(1)},...,\mathbf{k}^{(i)};t}\bra{ \mathbf{l}^{(1)},...,\mathbf{l}^{(j)};t}~,
\end{eqnarray}
where we have dropped the index $\phi$ from the density matrix elements for notational convenience. Furthermore, we realize that the influence action in Eq.~(\ref{eq:IFR}) can actually be expressed in terms of path integrals over $\chi$:
\begin{eqnarray}
\label{eq:SIFasepofchi}
    \exp\left[\widehat{S}_{IF}(t)\right] &=& 
\langle
\exp\left[\mathrm{i}\widehat{S}_\chi(t) + \mathrm{i}\widehat{S}_{\phi\chi}(t)\right]\rangle_\chi
\end{eqnarray}
with the expectation value for any operator $\widehat{A}$ that depends on $\chi$ given by \cite{Calzetta2008,Burrage2018,Kading:2022hhc}
\begin{eqnarray} 
\label{eq:SIFasPath}
\langle \widehat{A}\,\rangle_\chi &:=& \int d\chi^{\pm}_t d\chi^{\pm}_0 \delta(\chi_t^+-\chi_t^-)\rho_\chi [\chi^{\pm}_0;0]
\int^{\chi^{\pm}_t}_{\chi^{\pm}_0} \mathcal{D}\chi^{\pm} A[\chi^{a}]\exp \left\{ \mathrm{i}\widehat{S}[\chi;t] \right\}~,
\end{eqnarray} 
where $\rho_\chi[\chi^\pm_0;0] := \langle\chi^+_0|\hat{\rho}_\chi(0)|\chi^-_0\rangle$ is a density functional in the field basis; $A[\chi^{a}]$ is a projection of $\widehat{A}$ into the field basis, which in our case corresponds to the usual action functionals of $\chi$; and $\widehat{S}[\chi;t] := S[\chi^+;t] - S[\chi^-;t]$ is the free action functional of $\chi$. For practical purposes, we again focus on the perturbative case, such that  
\begin{eqnarray}
\label{eq:SIFasepofchiPer}
    \widehat{S}_{IF}(t) 
&\approx& 
\langle\widehat{S}_\chi(t) + \widehat{S}_{\phi\chi}(t) +\frac{\mathrm{i}}{2}
\Big[ 
\widehat{S}^2_\chi(t) + 2 \widehat{S}_\chi(t)\widehat{S}_{\phi\chi}(t) + \widehat{S}^2_{\phi\chi}(t)
\Big]
\rangle_\chi
\nonumber
\\
&&
- \frac{\mathrm{i}}{2}
\left[
\langle \widehat{S}_\chi(t) \rangle_\chi^2 
+2 \langle \widehat{S}_\chi(t) \rangle_\chi
\langle \widehat{S}_{\phi\chi}(t) \rangle_\chi
+
\langle \widehat{S}_{\phi\chi}(t) \rangle_\chi^2
\right]~.
\end{eqnarray}
Next, we define creators and annihilators of $\phi$ in the same way as in Eq.~(\ref{eq:CreatAnni}) and essentially follow the same procedure as in Sec.~\ref{sec:MomBasisClos} but now applied to Eq.~(\ref{eq:OQSPerturb}). Doing so, leads us to 
\begin{eqnarray}\label{eq:GenDensForm}
&&\rho_{g;h}(\mathbf{k}^{(1)},...,\mathbf{k}^{(g)};\mathbf{l}^{(1)},...,\mathbf{l}^{(h)};t)
= 
\nonumber
\\
&&
~~\sum\limits_{i,j=0}^\infty \frac{\mathrm{i}^{g+j}(-\mathrm{i})^{h+i}}{i!j!}
\lim_{\substack{x_{(1)}^{0},...,x_{(g)}^{0},x_{(1)}^{0\prime},...,x_{(h)}^{0\prime}\,\to\, t^{+}\\y_{(1)}^{0},...,y_{(i)}^{0},y_{(1)}^{0\prime},...,y_{(j)}^{0\prime}\,\to\, 0^-}}
\int \left(   \prod\limits_{a = 1}^id\Pi_{\mathbf{r}^{(a)}}\right)\left(\prod\limits_{b = 1}^j   d\Pi_{\mathbf{s}^{(b)}} \right) 
\nonumber
\\
&&
~~~~\times 
\rho_{i;j}(\mathbf{r}^{(1)},...,\mathbf{r}^{(i)};\mathbf{s}^{(1)},...,\mathbf{s}^{(j)};0) 
\int_{\mathbf{x}_{(1)}...\mathbf{x}_{(g)}\mathbf{x}_{(1)}^{\prime}...\mathbf{x}_{(h)}^{\prime}\mathbf{y}_{(1)}...\mathbf{y}_{(i)}\mathbf{y}_{(1)}^{\prime}...\mathbf{y}_{(j)}^{\prime}}
\nonumber
\\
&&
~~~~~~~\times 
\exp\left\{-\mathrm{i}\Bigg(\sum\limits^g_{a=1}\mathbf{k}^{(a)}\mathbf{x}_{(a)}
-\sum\limits^h_{a=1}\mathbf{l}^{(a)}\mathbf{x}_{(a)}^{\prime}\Bigg)
+ \mathrm{i}\Bigg(\sum\limits^i_{a=1}\mathbf{r}^{(a)} \mathbf{y}_{(a)}-\sum\limits^j_{a=1}\mathbf{s}^{(a)} \mathbf{y}_{(a)}^{\prime}\Bigg)\right\}
\nonumber
\\
&&
~~~~~~~\times 
\left(   \prod\limits_{a = 1}^g
\partial_{x^0_{(a)},E^\phi_{\mathbf{k}^{(a)}}} \right)
\left(  \prod\limits_{b = 1}^h
\partial_{x^{0\prime}_{(b)},E^\phi_{\mathbf{l}^{(b)}}}^*\right)
\left(\prod\limits_{c = 1}^i \partial_{y^0_{(c)},E^\phi_{\mathbf{r}^{(c)}}}^* \right)
\left(\prod\limits_{d = 1}^j 
\partial_{y^{0\prime}_{(d)},E^\phi_{\mathbf{s}^{(d)}}} 
\right)
\nonumber
\\
&&
~~~~~~~
\times 
{}_\phi\bbra{0} 
    \hat{\phi}^+_{x_{(1)}}...\hat{\phi}^+_{x_{(g)}}\hat{\phi}^-_{x^\prime_{(1)}}...\hat{\phi}^-_{x^\prime_{(h)}}
  e^{\mathrm{i}\widehat{S}_\phi(t) + \mathrm{i}\widehat{S}_{IF}(t)}
  \hat{\phi}^+_{y_{(1)}}...\hat{\phi}^+_{y_{(i)}}\hat{\phi}^-_{y^\prime_{(1)}}...\hat{\phi}^-_{y^\prime_{(j)}}
  \kket{0}_\phi
~.~~~~~
\end{eqnarray}
In the same way as we did for the total density matrix elements in Sec.~\ref{sec:MomBasisClos}, we translate the correlation function in Eq.~(\ref{eq:GenDensForm}) into a path integral, such that we obtain
\begin{eqnarray}\label{eq:OpenPaths}
&&\rho_{g;h}(\mathbf{k}^{(1)},...,\mathbf{k}^{(g)};\mathbf{l}^{(1)},...,\mathbf{l}^{(h)};t)
\nonumber
\\
&=&
~~\sum\limits_{i,j=0}^\infty \frac{\mathrm{i}^{g+j}(-\mathrm{i})^{h+i}}{i!j!}
\lim_{\substack{x_{(1)}^{0},...,x_{(g)}^{0},x_{(1)}^{0\prime},...,x_{(h)}^{0\prime}\,\to\, t^{+}\\y_{(1)}^{0},...,y_{(i)}^{0},y_{(1)}^{0\prime},...,y_{(j)}^{0\prime}\,\to\, 0^-}}
\int \left(   \prod\limits_{a = 1}^id\Pi_{\mathbf{r}^{(a)}}\right)\left(\prod\limits_{b = 1}^j   d\Pi_{\mathbf{s}^{(b)}} \right) 
\nonumber
\\
&&
~~~~\times 
\rho_{i;j}(\mathbf{r}^{(1)},...,\mathbf{r}^{(i)};\mathbf{s}^{(1)},...,\mathbf{s}^{(j)};0) 
\int_{\mathbf{x}_{(1)}...\mathbf{x}_{(g)}\mathbf{x}_{(1)}^{\prime}...\mathbf{x}_{(h)}^{\prime}\mathbf{y}_{(1)}...\mathbf{y}_{(i)}\mathbf{y}_{(1)}^{\prime}...\mathbf{y}_{(j)}^{\prime}}
\nonumber
\\
&&
~~~~~~~\times 
\exp\left\{-\mathrm{i}\Bigg(\sum\limits^g_{a=1}\mathbf{k}^{(a)}\mathbf{x}_{(a)}
-\sum\limits^h_{a=1}\mathbf{l}^{(a)}\mathbf{x}_{(a)}^{\prime}\Bigg)
+ \mathrm{i}\Bigg(\sum\limits^i_{a=1}\mathbf{r}^{(a)} \mathbf{y}_{(a)}-\sum\limits^j_{a=1}\mathbf{s}^{(a)} \mathbf{y}_{(a)}^{\prime}\Bigg)\right\}
\nonumber
\\
&&
~~~~~~~\times 
\left(   \prod\limits_{a = 1}^g
\partial_{x^0_{(a)},E^\phi_{\mathbf{k}^{(a)}}} \right)
\left(  \prod\limits_{b = 1}^h
\partial_{x^{0\prime}_{(b)},E^\phi_{\mathbf{l}^{(b)}}}^*\right)
\left(\prod\limits_{c = 1}^i \partial_{y^0_{(c)},E^\phi_{\mathbf{r}^{(c)}}}^* \right)
\left(\prod\limits_{d = 1}^j 
\partial_{y^{0\prime}_{(d)},E^\phi_{\mathbf{s}^{(d)}}} 
\right)
\nonumber
\\
&&
~~~~~~~\times 
\int\mathcal{D}\phi^{\pm} e^{\mathrm{i}\widehat{S}[\phi]}
\phi^+_{x_{(1)}}...\phi^+_{x_{(g)}}\phi^-_{x^\prime_{(1)}}...\phi^-_{x^\prime_{(h)}}
e^{\mathrm{i}\widehat{S}_\phi[t] + \mathrm{i}\widehat{S}_{IF}[\phi;t]}
\phi^+_{y_{(1)}}...\phi^+_{y_{(i)}}\phi^-_{y^\prime_{(1)}}...\phi^-_{y^\prime_{(j)}}
\nonumber
\\
&\approx&
~~\sum\limits_{i,j=0}^\infty \frac{\mathrm{i}^{g+j}(-\mathrm{i})^{h+i}}{i!j!}
\lim_{\substack{x_{(1)}^{0},...,x_{(g)}^{0},x_{(1)}^{0\prime},...,x_{(h)}^{0\prime}\,\to\, t^{+}\\y_{(1)}^{0},...,y_{(i)}^{0},y_{(1)}^{0\prime},...,y_{(j)}^{0\prime}\,\to\, 0^-}}
\int \left(   \prod\limits_{a = 1}^id\Pi_{\mathbf{r}^{(a)}}\right)\left(\prod\limits_{b = 1}^j   d\Pi_{\mathbf{s}^{(b)}} \right) 
\nonumber
\\
&&
~~~~\times 
\rho_{i;j}(\mathbf{r}^{(1)},...,\mathbf{r}^{(i)};\mathbf{s}^{(1)},...,\mathbf{s}^{(j)};0) 
\int_{\mathbf{x}_{(1)}...\mathbf{x}_{(g)}\mathbf{x}_{(1)}^{\prime}...\mathbf{x}_{(h)}^{\prime}\mathbf{y}_{(1)}...\mathbf{y}_{(i)}\mathbf{y}_{(1)}^{\prime}...\mathbf{y}_{(j)}^{\prime}}
\nonumber
\\
&&
~~~~~~~\times 
\exp\left\{-\mathrm{i}\Bigg(\sum\limits^g_{a=1}\mathbf{k}^{(a)}\mathbf{x}_{(a)}
-\sum\limits^h_{a=1}\mathbf{l}^{(a)}\mathbf{x}_{(a)}^{\prime}\Bigg)
+ \mathrm{i}\Bigg(\sum\limits^i_{a=1}\mathbf{r}^{(a)} \mathbf{y}_{(a)}-\sum\limits^j_{a=1}\mathbf{s}^{(a)} \mathbf{y}_{(a)}^{\prime}\Bigg)\right\}
\nonumber
\\
&&
~~~~~~~\times 
\left(   \prod\limits_{a = 1}^g
\partial_{x^0_{(a)},E^\phi_{\mathbf{k}^{(a)}}} \right)
\left(  \prod\limits_{b = 1}^h
\partial_{x^{0\prime}_{(b)},E^\phi_{\mathbf{l}^{(b)}}}^*\right)
\left(\prod\limits_{c = 1}^i \partial_{y^0_{(c)},E^\phi_{\mathbf{r}^{(c)}}}^* \right)
\left(\prod\limits_{d = 1}^j 
\partial_{y^{0\prime}_{(d)},E^\phi_{\mathbf{s}^{(d)}}} 
\right)
\nonumber
\\
&&
~~~~~~~\times 
\int\mathcal{D}\phi^{\pm} e^{\mathrm{i}\widehat{S}[\phi]}
\phi^+_{x_{(1)}}...\phi^+_{x_{(g)}}\phi^-_{x^\prime_{(1)}}...\phi^-_{x^\prime_{(h)}}
\bigg\{1 + \mathrm{i} \widehat{S}_\phi[\phi;t] + \mathrm{i}\langle\widehat{S}_\chi\rangle_\chi[\phi;t] + \mathrm{i}\langle\widehat{S}_{\phi\chi}\rangle_\chi[\phi;t] 
\nonumber
\\
&&~~~~~~~
  -  \frac{1}{2}\bigg[
\widehat{S}^2_\phi[\phi;t] + \langle\widehat{S}_\chi^2\rangle_\chi[\phi;t] + \langle\widehat{S}^2_{\phi\chi}\rangle_\chi[\phi;t]
+ 2 \widehat{S}_\phi[\phi;t]\langle\widehat{S}_\chi\rangle_\chi[\phi;t]
+ 2\widehat{S}_\phi[\phi;t]\langle\widehat{S}_{\phi\chi}\rangle_\chi[\phi;t]
 \nonumber
\\
&&~~~~~~~
 + 2\langle\widehat{S}_\chi\widehat{S}_{\phi\chi}\rangle_\chi[\phi;t] 
 \bigg]
  \bigg\}
\phi^+_{y_{(1)}}...\phi^+_{y_{(i)}}\phi^-_{y^\prime_{(1)}}...\phi^-_{y^\prime_{(j)}}\>,
\end{eqnarray}
where $\widehat{S}_{IF}[\phi;t]$ is the influence action functional of $\phi$ defined as the projection of the operator in Eq.~(\ref{eq:SIFasepofchi}) into the $\phi$-field space.
Note that, in contrast to the path integrals in Eq.~(\ref{eq:CloPath}), Eq.~(\ref{eq:SIFasPath}) also permits contractions between $\chi^+$ and $\chi^-$, which results in Wightman propagators \cite{Kading:2022jjl}. However, system degrees of freedoms in Eq.~(\ref{eq:OpenPaths}), i.e.,~$\phi$ in our case, still follow the rules outlined below Eq.~(\ref{eq:CloPath}) and consequently only allow for contractions of two equally-labelled fields.


\subsection{Master equations}

Finally, we want to show that Ref.~\cite{Kading:2022jjl} indeed presents general solutions to the master equation in Ref.~\cite{Burrage2018}. For this, we differentiate Eq.~(\ref{eq:Sol2cReduMP}) using the identities given in Eq.~(\ref{eq:TDTOP}) and obtain
\begin{eqnarray}
\label{eq:Master}
\partial_t\hat{\rho}_\mathcal{S}^+(t)\kket{1}_\mathcal{S}
&=&\mathrm{i}\partial_t[\widehat{S}_\mathcal{S}(t) + \widehat{S}_{IF}(t)]\exp\left[\mathrm{i}\widehat{S}_\mathcal{S}(t) + \mathrm{i}\widehat{S}_{IF}(t)\right]\hat{\rho}_\mathcal{S}^+(0)\kket{1}_\mathcal{S}
\nonumber
\\
&=&\mathrm{i}\partial_t[\widehat{S}_\mathcal{S}(t) + \widehat{S}_{IF}(t)]\hat{\rho}_\mathcal{S}^+(t)\kket{1}_\mathcal{S}
~.
\end{eqnarray}
Next, we restrict our discussion to second order in perturbation theory (as in Refs.~\cite{Burrage2018} and \cite{Kading:2022jjl}) and consequently work with Eq.~(\ref{eq:IFPerturbed}).
The master equation has the form
\begin{eqnarray}
\label{eq:mastereqgeneral}
\partial_t\hat{\rho}_\mathcal{S}^+(t)\kket{1}_\mathcal{S}
&\approx&
\bigg[ 
\mathrm{i}\partial_t(\widehat{S}_\mathcal{S}(t) + \braket{\widehat{S}_\mathcal{E}(t)} + \braket{\widehat{S}_{\mathcal{SE}}(t)})
- \braket{\widehat{S}_\mathcal{E}(t)\partial_t\widehat{S}_\mathcal{E}(t)} - \braket{\partial_t\widehat{S}_\mathcal{E}(t)\widehat{S}_{\mathcal{SE}}(t)}  
\nonumber
\\
&&
- \braket{\widehat{S}_\mathcal{E}(t)\partial_t\widehat{S}_{\mathcal{SE}}(t)} - \braket{\widehat{S}_{\mathcal{SE}}(t)\partial_t\widehat{S}_{\mathcal{SE}}(t)} + \braket{\widehat{S}_\mathcal{E}(t)}\braket{\partial_t\widehat{S}_\mathcal{E}(t)} 
+ \braket{\partial_t\widehat{S}_\mathcal{E}(t)}\braket{\widehat{S}_{\mathcal{SE}}(t)} \nonumber
\\
&&+ \braket{\widehat{S}_\mathcal{E}(t)}\braket{\partial_t\widehat{S}_{\mathcal{SE}}(t)}
+
\braket{\widehat{S}_{\mathcal{SE}}(t)}\braket{\partial_t\widehat{S}_{\mathcal{SE}}(t)}
\bigg] 
\hat{\rho}_\mathcal{S}^+(t)\kket{1}_\mathcal{S}
~.~~~~~
\end{eqnarray}
Considering our concrete case at hand, i.e.,~working with system $\phi$ and environment $\chi$, we can simplify Eq.~(\ref{eq:mastereqgeneral}),
\begin{eqnarray}
\label{eq:mastereqgeneral2}
\partial_t\hat{\rho}_\phi^+(t)\kket{1}_\phi
&\approx&
\bigg[ 
\mathrm{i}\partial_t(\widehat{S}_\phi(t) + \braket{\widehat{S}_{\mathcal{\phi\chi}}(t)})
- \braket{\partial_t\widehat{S}_\mathcal{\chi}(t)\widehat{S}_{\mathcal{\phi\chi}}(t)}  
- \braket{\widehat{S}_\mathcal{\chi}(t)\partial_t\widehat{S}_{\mathcal{\phi\chi}}(t)} - \braket{\widehat{S}_{\phi\chi}(t)\partial_t\widehat{S}_{\phi\chi}(t)} 
\nonumber
\\
&&+
\braket{\widehat{S}_{\mathcal{\phi\chi}}(t)}\braket{\partial_t\widehat{S}_{\mathcal{\phi\chi}}(t)}
\bigg] 
\hat{\rho}_\mathcal{S}^+(t)\kket{1}_\phi
~,
\end{eqnarray}
since we know that purely $\chi$-dependent terms have to vanish under the expectation value with respect to the environment \cite{Burrage2018,Kading:2022jjl} because they either produce tadpole diagrams, which cancel each other due to the structure of the doubled Hilbert space in TFD, or give rise to sums of correlation functions that complete the largest time equation \cite{tHooft:1973wag,Kobes:1985kc}.
Translating Eq.~(\ref{eq:mastereqgeneral2}) into a path integral expression for the reduced density matrix elements as in Sec.~\ref{sec:MomFockOpen}, we find  
\begin{eqnarray}
\label{eq:mastereqfinal}
&&\partial_t\rho_{g;h}(\mathbf{k}^{(1)},...,\mathbf{k}^{(g)};\mathbf{l}^{(1)},...,\mathbf{l}^{(h)};t)
\nonumber
\\
&=&
-\mathrm{i}\left(E^\phi_{\mathbf{k}^{(1)}} +...+E^\phi_{\mathbf{k}^{(g)}} - E^\phi_{\mathbf{l}^{(1)}}-...-E^\phi_{\mathbf{l}^{(h)}} \right)\rho_{g;h}(\mathbf{k}^{(1)},...,\mathbf{k}^{(g)};\mathbf{l}^{(1)},...,\mathbf{l}^{(h)};t)
\nonumber
\\
&&
\quad
+\sum\limits_{i,j=0}^\infty \frac{\mathrm{i}^{g+j}(-\mathrm{i})^{h+i}}{i!j!}
\lim_{\substack{x_{(1)}^{0},...,x_{(g)}^{0},x_{(1)}^{0\prime},...,x_{(h)}^{0\prime}\,\to\, t^{+}\\y_{(1)}^{0},...,y_{(i)}^{0},y_{(1)}^{0\prime},...,y_{(j)}^{0\prime}\,\to\, 0^-}}
\int \left(   \prod\limits_{a = 1}^id\Pi_{\mathbf{r}^{(a)}} 
e^{\mathrm{i}E^\phi_{\mathbf{r}^{(a)}}t}
\right)
\left(\prod\limits_{b = 1}^j   d\Pi_{\mathbf{s}^{(b)}} 
e^{-\mathrm{i}E^\phi_{\mathbf{s}^{(b)}}t}
\right) 
\nonumber
\\
&&
\quad\times
\rho_{i;j}(\mathbf{r}^{(1)},...,\mathbf{r}^{(i)};\mathbf{s}^{(1)},...,\mathbf{s}^{(j)};t) 
\int_{\mathbf{x}_{(1)}...\mathbf{x}_{(g)}\mathbf{x}_{(1)}^{\prime}...\mathbf{x}_{(h)}^{\prime}\mathbf{y}_{(1)}...\mathbf{y}_{(i)}\mathbf{y}_{(1)}^{\prime}...\mathbf{y}_{(j)}^{\prime}}
\nonumber
\\
&&
\quad\times 
\exp\left\{-\mathrm{i}\Bigg(\sum\limits^g_{a=1}\mathbf{k}^{(a)}\mathbf{x}_{(a)}
-\sum\limits^h_{a=1}\mathbf{l}^{(a)}\mathbf{x}_{(a)}^{\prime}\Bigg)
+ \mathrm{i}\Bigg(\sum\limits^i_{a=1}\mathbf{r}^{(a)} \mathbf{y}_{(a)}-\sum\limits^j_{a=1}\mathbf{s}^{(a)} \mathbf{y}_{(a)}^{\prime}\Bigg)\right\}
\nonumber
\\
&&
\quad\times 
\left(   \prod\limits_{a = 1}^g
\partial_{x^0_{(a)},E^\phi_{\mathbf{k}^{(a)}}} \right)
\left(  \prod\limits_{b = 1}^h
\partial_{x^{0\prime}_{(b)},E^\phi_{\mathbf{l}^{(b)}}}^*\right)
\left(\prod\limits_{c = 1}^i \partial_{y^0_{(c)},E^\phi_{\mathbf{r}^{(c)}}}^* \right)
\left(\prod\limits_{d = 1}^j 
\partial_{y^{0\prime}_{(d)},E^\phi_{\mathbf{s}^{(d)}}} 
\right)
\nonumber
\\
&&
\quad\times 
\int\mathcal{D}\phi^{\pm} e^{\mathrm{i}\widehat{S}[\phi]}
\phi^+_{x_{(1)}}...\phi^+_{x_{(g)}}\phi^-_{x^\prime_{(1)}}...\phi^-_{x^\prime_{(h)}}
\bigg[ 
\mathrm{i}\partial_t(\widehat{S}_\phi[\phi;t] + \widehat{S}_{IF}[\phi;t])
\bigg] 
\phi^+_{y_{(1)}}...\phi^+_{y_{(i)}}\phi^-_{y^\prime_{(1)}}...\phi^-_{y^\prime_{(j)}}
\nonumber\\
&\approx&
-\mathrm{i}\left(E^\phi_{\mathbf{k}^{(1)}} +...+E^\phi_{\mathbf{k}^{(g)}} - E^\phi_{\mathbf{l}^{(1)}}-...-E^\phi_{\mathbf{l}^{(h)}} \right)\rho_{g;h}(\mathbf{k}^{(1)},...,\mathbf{k}^{(g)};\mathbf{l}^{(1)},...,\mathbf{l}^{(h)};t)
\nonumber
\\
&&
\quad
+\sum\limits_{i,j=0}^\infty \frac{\mathrm{i}^{g+j}(-\mathrm{i})^{h+i}}{i!j!}
\lim_{\substack{x_{(1)}^{0},...,x_{(g)}^{0},x_{(1)}^{0\prime},...,x_{(h)}^{0\prime}\,\to\, t^{+}\\y_{(1)}^{0},...,y_{(i)}^{0},y_{(1)}^{0\prime},...,y_{(j)}^{0\prime}\,\to\, 0^-}}
\int \left(   \prod\limits_{a = 1}^id\Pi_{\mathbf{r}^{(a)}} 
e^{\mathrm{i}E^\phi_{\mathbf{r}^{(a)}}t}
\right)
\left(\prod\limits_{b = 1}^j   d\Pi_{\mathbf{s}^{(b)}} 
e^{-\mathrm{i}E^\phi_{\mathbf{s}^{(b)}}t}
\right) 
\nonumber
\\
&&
\quad\times
\rho_{i;j}(\mathbf{r}^{(1)},...,\mathbf{r}^{(i)};\mathbf{s}^{(1)},...,\mathbf{s}^{(j)};t) 
\int_{\mathbf{x}_{(1)}...\mathbf{x}_{(g)}\mathbf{x}_{(1)}^{\prime}...\mathbf{x}_{(h)}^{\prime}\mathbf{y}_{(1)}...\mathbf{y}_{(i)}\mathbf{y}_{(1)}^{\prime}...\mathbf{y}_{(j)}^{\prime}}
\nonumber
\\
&&
\quad\times 
\exp\left\{-\mathrm{i}\Bigg(\sum\limits^g_{a=1}\mathbf{k}^{(a)}\mathbf{x}_{(a)}
-\sum\limits^h_{a=1}\mathbf{l}^{(a)}\mathbf{x}_{(a)}^{\prime}\Bigg)
+ \mathrm{i}\Bigg(\sum\limits^i_{a=1}\mathbf{r}^{(a)} \mathbf{y}_{(a)}-\sum\limits^j_{a=1}\mathbf{s}^{(a)} \mathbf{y}_{(a)}^{\prime}\Bigg)\right\}
\nonumber
\\
&&
\quad\times 
\left(   \prod\limits_{a = 1}^g
\partial_{x^0_{(a)},E^\phi_{\mathbf{k}^{(a)}}} \right)
\left(  \prod\limits_{b = 1}^h
\partial_{x^{0\prime}_{(b)},E^\phi_{\mathbf{l}^{(b)}}}^*\right)
\left(\prod\limits_{c = 1}^i \partial_{y^0_{(c)},E^\phi_{\mathbf{r}^{(c)}}}^* \right)
\left(\prod\limits_{d = 1}^j 
\partial_{y^{0\prime}_{(d)},E^\phi_{\mathbf{s}^{(d)}}} 
\right)
\nonumber
\\
&&
\quad\times 
\int\mathcal{D}\phi^{\pm} e^{\mathrm{i}\widehat{S}[\phi]}
\phi^+_{x_{(1)}}...\phi^+_{x_{(g)}}\phi^-_{x^\prime_{(1)}}...\phi^-_{x^\prime_{(h)}}
\bigg[ 
\mathrm{i}\partial_t(\widehat{S}_\phi[\phi;t] + \braket{\widehat{S}_{\mathcal{\phi\chi}}}[\phi;t])
- \braket{\partial_t\widehat{S}_\mathcal{\chi}\widehat{S}_{\mathcal{\phi\chi}}}[\phi;t]  
\nonumber
\\
&&
\quad- \braket{\widehat{S}_\mathcal{\chi}\partial_t\widehat{S}_{\mathcal{\phi\chi}}}[\phi;t] - \braket{\widehat{S}_{\phi\chi}\partial_t\widehat{S}_{\phi\chi}}[\phi;t] +
\braket{\widehat{S}_{\mathcal{\phi\chi}}}[\phi;t]\braket{\partial_t\widehat{S}_{\mathcal{\phi\chi}}}[\phi;t]
\bigg] 
\phi^+_{y_{(1)}}...\phi^+_{y_{(i)}}\phi^-_{y^\prime_{(1)}}...\phi^-_{y^\prime_{(j)}}~,~~~
\end{eqnarray}
where the first term on the right-hand side stems from projecting the time derivative on the left-hand side of Eq.~(\ref{eq:mastereqgeneral2}) into the Fock basis. Setting $g=h=1$, assuming that only the density matrix elements for $i=j=1$ on the right-hand side of Eq.~(\ref{eq:mastereqfinal}) are non-vanishing, and considering the case of no self-interactions for $\phi$, we recover the master equation presented in Ref.~\cite{Burrage2018}, up to the purely environment-dependent terms that we have already dropped in Eq.~(\ref{eq:mastereqgeneral2}) but which also vanish at a later point in Ref.~\cite{Burrage2018}. This means that we have just shown that, as anticipated, the method presented in Ref.~\cite{Kading:2022jjl} must solve the master equation from Ref.~\cite{Burrage2018}. Consequently, since in Sec.~\ref{sec:Open} we have demonstrated that the approach from Ref.~\cite{Kading:2022jjl} does indeed describe non-Markovian dynamics, we can confirm the claim of Ref.~\cite{Burrage2018} that their master equation is also non-Markovian. In addition, Eq.~(\ref{eq:mastereqfinal}) is a generalization of the method presented in Ref.~\cite{Burrage2018} to arbitrary occupation numbers in Fock space. 

It is also worth noting that both the general master equation (\ref{eq:Master}) and the specific master equation (\ref{eq:mastereqfinal}) only contain density matrix elements at the final time $t$ and no time integrals that include the density matrix elements. While this might, at first glance, seem unusual since more frequently used master equations like the Lindblad equation \cite{Manzano_2020} contain such time integrals, this is not as surprising if we consider that the quantum Liouville equation (\ref{eq:LiouvilleInt}) does not contain such integrals as well. In fact, since we know that the solutions in Eqs.~(\ref{eq:Sol1}) and (\ref{eq:Sol2}) are equivalent, we can clearly see that such time integrals are to some extent artificial in the sense that master equations containing them, e.g., Lindblad \cite{Manzano_2020}, are usually derived by using Eq.~(\ref{eq:Sol1}). Master equations that are instead derived from Eq.~(\ref{eq:Sol2}) should generally not contain any time integrals over density matrix elements and only density matrix elements at one particular time.


\section{Conclusion}
\label{sec:Conclusion}

Density matrices are invaluable mathematical tools for the description of closed or open quantum systems. In recent years, powerful formalisms for computing density matrix elements in quantum field theory, based on TFD and the Schwinger-Keldysh formalism, have been developed: Ref.~\cite{Burrage2018} derived a formula for quantum master equations, and Refs.~\cite{Kading:2022jjl} and \cite{Kading:2022hhc} discussed ways of directly computing density matrix elements for open and closed systems, respectively.

In this article, we provided a detailed discussion of the methods presented in Refs.~\cite{Burrage2018,Kading:2022jjl,Kading:2022hhc} in order to make them accessible for interested researchers, especially from the open quantum system and relativistic field theory communities. We used this opportunity to also address some previously unanswered questions regarding these methods and open quantum systems in general. 

At first, we discussed general exact and perturbative solutions of the quantum Liouville equation in the framework of TFD. We then demonstrated that dynamical maps describing the dynamics of general closed systems are indeed divisible. Subsequently, we focused on a closed system comprising two interacting real scalar field species and derived a path integral-based formula for directly computing the total system's density matrix elements in Fock space. This was complemented by a discussion of situations in which the found formula is particularly applicable. Next, we moved on to a discussion of open quantum systems. We derived the influence action operator in an exact form and up to second order in perturbation theory. Afterwards, we showed that, without any approximations or assumptions, even the exact dynamical maps of general open quantum systems are not divisible, which implies that they are describing non-Markovian dynamics. In this way, we have made a topical observation about general open quantum systems and verified the claims of Refs.~\cite{Burrage2018,Kading:2022jjl} that their formalisms describe non-Markovian dynamics. After using the Born approximation in the perturbative case, we found that divisibility is broken by connected correlation functions of the environment and the system-environment interaction. Studying these divisibility-breaking terms in a future work might be a worthwhile endeavour and make a contribution to the ongoing discussions about quantum non-Markovianity. After deriving path integral-based expressions for directly computing the reduced density matrix elements of an open quantum system constituted by the scalar field $\phi$ interacting with an environmental scalar field $\chi$, we showed how the results from Ref.~\cite{Kading:2022jjl} are related to those from Ref.~\cite{Burrage2018}, and obtained a generalization to arbitrary Fock space states of the master equation formalism given in Ref.~\cite{Burrage2018}. Finally, we discussed why the master equations presented in this article do not contain time integrals over the density matrix elements and explained that such integrals are somewhat artificial since they stem from choosing a particular solution of the quantum Liouville equation.


\begin{acknowledgments}
The authors are grateful to N.~Agarwal, B.~Bowen, A.~Kamal, and P.~Millington for helpful discussions.
This research was funded in whole or in part by the Austrian Science Fund (FWF) [10.55776/PAT8564023] and [10.55776/PAT7599423]. For open access purposes, the author has applied a CC BY public copyright license to any author accepted manuscript version arising from this submission.
\end{acknowledgments}


\appendix

\section{Schr\"odinger-like form of the quantum Liouville equation}
\label{AppQL}

Here, we show that the quantum Liouville equation (\ref{eq:LiouvilleInt}) can be brought into a Schr\"odinger-like form (\ref{eq:SchrLiouvEq}) in TFD. After promoting the operators in Eq.~(\ref{eq:LiouvilleInt}) to $+$-labeled operators in the doubled Hilbert space, see Eq.~(\ref{eq:TFDoper}), we find 
\begin{eqnarray}
    \frac{\partial}{\partial t} \hat{\rho}^+(t) \kket{1} &=& -\mathrm{i}\{[\hat{H}_I(t)\hat{\rho}(t) - \hat{\rho}(t)\hat{H}_I(t)] \otimes \hat{\mathbb{I}}\}  \kket{1}~. 
\end{eqnarray}
Comparing this with Eq.~(\ref{eq:SchrLiouvEq}), we conclude that we only need to show that 
\begin{eqnarray}
\label{eq:AppLRsame}
    [\hat{H}_I(t) \otimes \hat{\mathbb{I}}]  \kket{1} &=&      [\hat{\mathbb{I}} \otimes \hat{H}_I(t)]  \kket{1}
\end{eqnarray}
is true in order for Eq.~(\ref{eq:SchrLiouvEq}) to indeed represent a Schr\"odinger-like form of the quantum Liouville equation (\ref{eq:LiouvilleInt}). In order to prove that Eq.~(\ref{eq:AppLRsame}) is indeed fulfilled, we expand state $\kket{1}$ in terms of eigenfunctions $\ket{\mathfrak{n}}$ of the Hamiltonian
\begin{eqnarray}
\kket{1} &=& \sumint_{\mathfrak{n}}\ket{\mathfrak{n}}\otimes\ket{\mathfrak{n}}\,\,\,,
\end{eqnarray}
which is always possible and find
\begin{eqnarray}
  [\hat{H}_I(t) \otimes \hat{\mathbb{I}}]  \kket{1}  &=& [\hat{H}_I(t) \otimes \hat{\mathbb{I}}]\sumint_{\mathfrak{n}}\ket{\mathfrak{n}}\otimes\ket{\mathfrak{n}}
  \nonumber\\
&=& \sumint_{\mathfrak{n}}\hat{H}_I(t)\ket{\mathfrak{n}}\otimes\ket{\mathfrak{n}}
  \nonumber\\
  &=& \sumint_{\mathfrak{n}}E_{\mathfrak{n}}\ket{\mathfrak{n}}\otimes\ket{\mathfrak{n}}
  \nonumber\\
  &=& \sumint_{\mathfrak{n}}\ket{\mathfrak{n}}\otimes E_{\mathfrak{n}}\ket{\mathfrak{n}}
  \nonumber\\
  &=& [\hat{\mathbb{I}} \otimes  \hat{H}_I(t)]\sumint_{\mathfrak{n}}\ket{\mathfrak{n}}\otimes\ket{\mathfrak{n}}\,\,\,.
\end{eqnarray}
Consequently, we now know that Eq.~(\ref{eq:AppLRsame}) is fulfilled. In conclusion, this proves that Eq.~(\ref{eq:SchrLiouvEq}) is indeed a Schrödinger-like form of the quantum Liouville equation (\ref{eq:LiouvilleInt}).


\section{Derivatives of exponential operators}
\label{AppA}

Here, we follow Ref.~\cite{MarioHabil} in order to prove Eq.~(\ref{eq:TDTOP}), i.e.,
\begin{eqnarray}
 \frac{d}{dt'}\exp\left[{\mathrm{i}\widehat{S}_I(t,t')}\right]&=&-\mathrm{i}\exp\left[{\mathrm{i}\widehat{S}_I(t,t')}\right]\frac{d}{dt'} \widehat{S}_I(t') ~,~~~
 \frac{d}{dt'}\exp\left[{\mathrm{i}\widehat{S}_I(t')}\right]\,=\,\mathrm{i}\frac{d}{dt'} \widehat{S}_I(t')\exp\left[{\mathrm{i}\widehat{S}_I(t')}\right]~.~~~~~~~
\end{eqnarray}
For this, we return to the Hamiltonian description that we have used before introducing action operators in Eq.~(\ref{eq:DefAction}). We begin with
\begin{eqnarray} 
\label{eq:AppDer1}
\frac{d}{dt'}\int_{t'}^tdt_1\cdots\int_{t'}^tdt_n\,\text{T}\!\left(\widehat{H}_I(t_1)\cdots\widehat{H}_I(t_n)\right) 
&=&
\lim_{\Delta\to0}\frac{1}{\Delta}\left[\int_{t'+\Delta}^{t}dt_1\cdots\int_{t'+\Delta}^{t}dt_n\,\text{T}\!\left(\widehat{H}_I(t_1)\cdots\widehat{H}_I(t_n)\right) \right.
\nonumber
\\
&&
~~~~~~~~
\left.
- \int_{t'}^tdt_1\cdots\int_{t'}^tdt_n\,\text{T}\!\left(\widehat{H}_I(t_1)\cdots\widehat{H}_I(t_n)\right)\right]~.~~~~~~~
\end{eqnarray}
Next, we consider only the first term in the square bracket on the right-hand side of Eq.~(\ref{eq:AppDer1}):
\begin{eqnarray}
\label{eq:AppDer2}
&&\int_{t'+\Delta}^{t}dt_1\cdots\int_{t'+\Delta}^{t}dt_n\,\text{T}\!\left(\widehat{H}_I(t_1)\cdots\widehat{H}_I(t_n)\right) 
\nonumber
\\
&=&
\int_{t'}^tdt_1\int_{t'+\Delta}^{t}dt_2\cdots\int_{t'+\Delta}^{t}dt_n\,\text{T}\!\left(\widehat{H}_I(t_1)\cdots\widehat{H}_I(t_n)\right) 
\nonumber
\\
&&
~~~~~~~~~~~~~~~~~~~~
- \underbrace{\strut  \Delta\int_{t'+\Delta}^{t}dt_2\cdots\int_{t'+\Delta}^{t}dt_n\,\text{T}\!\left(\widehat{H}_I(t_2)\cdots\widehat{H}_I(t_n)\right)\widehat{H}_I(t')}_{\mathclap{\displaystyle\Delta\int_{t'}^tdt_2\cdots\int_{t'}^tdt_n\,\text{T}\!\left(\widehat{H}_I(t_2)\cdots\widehat{H}_I(t_n)\right)\widehat{H}_I(t')  + \mathcal O(\Delta^2)}} + \mathcal O(\Delta^2) 
\nonumber
\\
&=&
\int_{t'}^tdt_1\int_{t'}^tdt_2\int_{t'+\Delta}^{t}dt_3\cdots\int_{t'+\Delta}^{t}dt_n\,\text{T}\!\left(\widehat{H}_I(t_1)\cdots\widehat{H}_I(t_n)\right) 
\nonumber
\\
&&
~~~~~~~~~~~~~~~~~~~~
- \Delta\int_{t'}^tdt_1\int_{t'+\Delta}^{t}dt_3\cdots\int_{t'+\Delta}^{t}dt_n\,\text{T}\!\left(\widehat{H}_I(t_1)\widehat{H}_I(t_3)\cdots\widehat{H}_I(t_n)\right)\widehat{H}_I(t') 
\nonumber
\\
&&
~~~~~~~~~~~~~~~~~~~~
- \Delta\int_{t'}^tdt_2\cdots\int_{t'}^tdt_n\,\text{T}\!\left(\widehat{H}_I(t_2)\cdots\widehat{H}_I(t_n)\right)\widehat{H}_I(t') + \mathcal O(\Delta^2)  
\nonumber
\\
&=&
\int_{t'}^tdt_1\int_{t'}^tdt_2\int_{t'+\Delta}^{t}dt_3\cdots\int_{t'+\Delta}^{t}dt_n\,\text{T}\!\left(\widehat{H}_I(t_1)\cdots\widehat{H}_I(t_n)\right)
\nonumber
\\
&&
~~~~~~~~~~~~~~~~~~~~
- 2\Delta\int_{t'}^tdt_1\cdots\int_{t'}^tdt_{n-1}\,\text{T}\!\left(\widehat{H}_I(t_1)\cdots\widehat{H}_I(t_{n-1})\right)\widehat{H}_I(t') + \mathcal O(\Delta^2)  
\nonumber
\\
&=&
\int_{t'}^tdt_1\cdots\int_{t'}^tdt_n\,\text{T}\!\left(\widehat{H}_I(t_1)\cdots\widehat{H}_I(t_n)\right)
\nonumber
\\
&&
~~~~~~~~~~~~~~~~~~~~
- n\Delta\int_{t'}^tdt_1\cdots\int_{t'}^tdt_{n-1}\,\text{T}\!\left(\widehat{H}_I(t_1)\cdots\widehat{H}_I(t_{n-1})\right)\widehat{H}_I(t') + \mathcal O(\Delta^2)~.
\end{eqnarray}
Substituting Eq.~(\ref{eq:AppDer2}) into Eq.~(\ref{eq:AppDer1}) gives us
\begin{eqnarray}
&&\frac{d}{dt'}\int_{t'}^tdt_1\cdots\int_{t'}^tdt_n\,\text{T}\!\left(\widehat{H}_I(t_1)\cdots\widehat{H}_I(t_n)\right) 
\nonumber
\\
&&
~~~~~~~~~~~~~~~~~~~~
\,=\,
-n\int_{t'}^tdt_1\cdots\int_{t'}^tdt_{n-1}\,\text{T}\!\left(\widehat{H}_I(t_1)\cdots\widehat{H}_I(t_{n-1})\right)\widehat{H}_I(t')~,
\end{eqnarray}
from which we conclude that
\begin{eqnarray}
\frac{d}{dt'}\,\text{T}\!\left(e^{\textstyle -\mathrm{i}\int_{t'}^td\tau\,\widehat{H}_I(\tau)}\right) 
&=&
\mathrm{i}\widehat{H}_I(t') - \frac{\mathrm{i}^2}{1!}\int_{t'}^td\tau\,\widehat{H}_I(\tau)\widehat{H}_I(t') + \ldots 
\nonumber
\\
&=&
\mathrm{i}\text{T}\!\left(e^{\textstyle -\mathrm{i}\int_{t'}^td\tau\,\widehat{H}_I(\tau)}\right)\widehat{H}_I(t')~.
\end{eqnarray}
Using Eq.~(\ref{eq:DefAction}), we finally obtain
\begin{eqnarray}
 \frac{d}{dt'}\exp\left[{\mathrm{i}\widehat{S}_I(t,t')}\right] 
 &=& 
 -\mathrm{i}\exp\left[{\mathrm{i}\widehat{S}_I(t,t')}\right]\frac{d}{dt'} \widehat{S}_I(t')~,
\end{eqnarray}
which proves the first equality in Eq.~(\ref{eq:TDTOP}).
Analogously, we use
\begin{eqnarray}
\frac{d}{dt'}\int_0^{t'}dt_1\cdots\int_0^{t'}dt_n\,\text{T}\!\left(\widehat{H}_I(t_1)\cdots\widehat{H}_I(t_n)\right) 
&=&
\lim_{\Delta\to0}\frac{1}{\Delta}\left[\int_0^{t'+\Delta}dt_1\cdots\int_0^{t'+\Delta}dt_n\,\text{T}\!\left(\widehat{H}_I(t_1)\cdots\widehat{H}_I(t_n)\right) 
\right.
\nonumber
\\
&&
~~~~~~~~
\left.
- \int_0^{t'}dt_1\cdots\int_0^{t'}dt_n\,\text{T}\!\left(\widehat{H}_I(t_1)\cdots\widehat{H}_I(t_n)\right)\right]~.~~~~~~~
\end{eqnarray}
Subsequently,
\begin{eqnarray}
&&
\int_0^{t'+\Delta}dt_1\cdots\int_0^{t'+\Delta}dt_n\,\text{T}\!\left(\widehat{H}_I(t_1)\cdots\widehat{H}_I(t_n)\right) 
\nonumber
\\
&=& 
\int_0^{t'}dt_1\int_0^{t'+\Delta}dt_2\cdots\int_0^{t'+\Delta}dt_n\,\text{T}\!\left(\widehat{H}_I(t_1)\cdots\widehat{H}_I(t_n)\right) 
\nonumber
\\
&&
~~~~~~~~~~~~~~~~~~~~
+ \underbrace{\strut  \Delta\widehat{H}_I(t')\int_0^{t'+\Delta}dt_2\cdots\int_0^{t'+\Delta}dt_n\,\text{T}\!\left(\widehat{H}_I(t_2)\cdots\widehat{H}_I(t_n)\right)}_{\mathclap{\displaystyle\Delta\widehat{H}_I(t')\int_0^{t'}dt_2\cdots\int_0^{t'}dt_n\,\text{T}\!\left(\widehat{H}_I(t_2)\cdots\widehat{H}_I(t_n)\right)  + \mathcal O(\Delta^2)}} + \mathcal O(\Delta^2) 
\nonumber
\\
&=&
\int_0^{t'}dt_1\int_0^{t'}dt_2\int_0^{t'+\Delta}dt_3\cdots\int_0^{t'+\Delta}dt_n\,\text{T}\!\left(\widehat{H}_I(t_1)\cdots\widehat{H}_I(t_n)\right)
\nonumber
\\
&&
~~~~~~~~~~~~~~~~~~~~
+ \Delta\widehat{H}_I(t')\int_0^{t'}dt_1\int_0^{t'+\Delta}dt_3\cdots\int_0^{t'+\Delta}dt_n\,\text{T}\!\left(\widehat{H}_I(t_1)\widehat{H}_I(t_3)\cdots\widehat{H}_I(t_n)\right) 
\nonumber
\\
&&
~~~~~~~~~~~~~~~~~~~~
+ \Delta\widehat{H}_I(t')\int_0^{t'}dt_2\cdots\int_0^{t'}dt_n\,\text{T}\!\left(\widehat{H}_I(t_2)\cdots\widehat{H}_I(t_n)\right) + \mathcal O(\Delta^2)  
\nonumber
\\
&=&
\int_0^{t'}dt_1\int_0^{t'}dt_2\int_0^{t'+\Delta}dt_3\cdots\int_0^{t'+\Delta}dt_n\,\text{T}\!\left(\widehat{H}_I(t_1)\cdots\widehat{H}_I(t_n)\right) 
\nonumber
\\
&&
~~~~~~~~~~~~~~~~~~~~
+ 2\Delta\widehat{H}_I(t')\int_0^{t'}dt_1\cdots\int_0^{t'}dt_{n-1}\,\text{T}\!\left(\widehat{H}_I(t_1)\cdots\widehat{H}_I(t_{n-1})\right) + \mathcal O(\Delta^2)  
\nonumber
\\
&=&
\int_0^{t'}dt_1\cdots\int_0^{t'}dt_n\,\text{T}\!\left(\widehat{H}_I(t_1)\cdots\widehat{H}_I(t_n)\right) 
\nonumber
\\
&&
~~~~~~~~~~~~~~~~~~~~
+ n\Delta\widehat{H}_I(t')\int_0^{t'}dt_1\cdots\int_0^{t'}dt_{n-1}\,\text{T}\!\left(\widehat{H}_I(t_1)\cdots\widehat{H}_I(t_{n-1})\right) + \mathcal O(\Delta^2)~,
\end{eqnarray}
which leads us to
\begin{eqnarray}
&&\frac{d}{dt'}\int_0^{t'}dt_1\cdots\int_0^{t'}dt_n\,\text{T}\!\left(\widehat{H}_I(t_1)\cdots\widehat{H}_I(t_n)\right)
\nonumber
\\
&&
~~~~~~~~~~~~~~~~~~~~
\,=\,
n\widehat{H}_I(t')\int_0^{t'}dt_1\cdots\int_0^{t'}dt_{n-1}\,\text{T}\!\left(\widehat{H}_I(t_1)\cdots\widehat{H}_I(t_{n-1})\right)~.
\end{eqnarray}
Therefore, we conclude
\begin{eqnarray}
\frac{d}{dt'}\,\text{T}\!\left(e^{\textstyle -\mathrm{i}\int_0^{t'}d\tau\,\widehat{H}_I(\tau)}\right) 
&=&
-\mathrm{i}\widehat{H}_I(t') + \frac{\mathrm{i}^2}{1!}\,\widehat{H}_I(t')\int_0^{t'}d\tau\,\widehat{H}_I(\tau) + \ldots 
\nonumber
\\
&=&
-\mathrm{i}\widehat{H}_I(t')\,\text{T}\!\left(e^{\textstyle -\mathrm{i}\int_0^{t'}d\tau\,\widehat{H}_I(\tau)}\right)~, 
\end{eqnarray}
and, using Eq.~(\ref{eq:DefAction}), prove that also
\begin{eqnarray}
 \frac{d}{dt'}\exp\left[{\mathrm{i}\widehat{S}_I(t')}\right]
 &=&
 \mathrm{i}\frac{d}{dt'} \widehat{S}_I(t')\exp\left[{\mathrm{i}\widehat{S}_I(t')}\right]
\end{eqnarray}
is true.


\bibliography{Bib}
\bibliographystyle{JHEP}

\end{document}